\documentclass[fleqn,usenatbib]{mnras}

\usepackage{newtxtext,newtxmath}

\usepackage[T1]{fontenc}

\DeclareRobustCommand{\VAN}[3]{#2}
\let\VANthebibliography\thebibliography
\def\thebibliography{\DeclareRobustCommand{\VAN}[3]{##3}\VANthebibliography}

\usepackage{graphicx}	
\usepackage{amsmath}	
\usepackage{xcolor}     
\usepackage{subcaption}
\usepackage{caption}

\title[High-resolution retrievals of MASCARA-1b]{High-resolution emission spectroscopy retrievals of MASCARA-1b with CRIRES+: Strong detections of CO, H$_2$O and Fe emission lines and a C$/$O consistent with solar}

\author[S. Ramkumar et al.]{
Swaetha Ramkumar$^{1}$\thanks{E-mail: ramkumas@tcd.ie},
Neale P. Gibson$^{1}$,
Stevanus K. Nugroho$^{2,3}$,
Cathal Maguire$^{1}$,
and Mark Fortune$^{1}$
\vspace{-0.3cm} \\ \\
$^{1}$School of Physics, Trinity College Dublin, University of Dublin, Dublin-2, Ireland\\
$^{2}$Astrobiology Center, NINS, 2-21-1 Osawa, Mitaka, Tokyo 181-8588, Japan\\
$^{3}$National Astronomical Observatory of Japan, 2-21-1 Osawa, Mitaka, Tokyo 181-8588, Japan}

\date{Accepted 2023 August 11. Received 2023 July 20; in original form 2023 June 20}

\pubyear{2023}

\begin{document}
\label{firstpage}
\pagerange{\pageref{firstpage}--\pageref{lastpage}}
\maketitle

\begin{abstract}
The characterization of exoplanet atmospheres has proven to be successful using high-resolution spectroscopy. Phase curve observations of hot/ultra-hot Jupiters can reveal their compositions and thermal structures, thereby allowing the detection of molecules and atoms in the planetary atmosphere using the cross-correlation technique. We present pre-eclipse observations of the ultra-hot Jupiter, MASCARA-1b, observed with the recently upgraded CRIRES+ high-resolution infrared spectrograph at the VLT. We report a detection of $\rm{Fe}$ ($\approx$\,8.3$\,\sigma$) in the K-band and confirm previous detections of $\rm{CO}$ ($ {>}15\,\sigma $) and $ \mathrm{H_2O} $ ($ {>}10\,\sigma $) in the day-side atmosphere of MASCARA-1b. Using a Bayesian inference framework, we retrieve the abundances of the detected species and constrain planetary orbital velocities, $T$-$P$ profiles, and the carbon-to-oxygen ratio ($\rm C/O$). A free retrieval results in an elevated $\mathrm{CO}$ abundance ($ \log_{10}(\chi_{\rm{{}^{12}CO}}) = -2.85^{+0.57}_{-0.69} $), leading to a super-solar $\rm C/O$ ratio. More realistically, allowing for vertically-varying chemistry in the atmosphere by incorporating a chemical-equilibrium model results in a $\rm C/O$ of $0.68^{+0.12}_{-0.22}$ and a metallicity of $[\rm M/H] = 0.62^{+0.28}_{-0.55}$, both consistent with solar values. Finally, we also report a slight offset of the $ \rm{Fe} $ feature in both $ K_{\rm p} $ and $ v_{\rm sys} $ that could be a signature of atmospheric dynamics. Due to the 3D structure of exoplanet atmospheres and the exclusion of time/phase dependence in our 1D forward models, further follow-up observations and analysis are required to confirm or refute this result.
\end{abstract}




\section{Introduction}
High-resolution Doppler-resolved spectroscopy ($\mathrm{R}$\,$\gtrsim$\,$25{,}000$) has shown to be extremely effective in characterizing exoplanet atmospheres \citep[e.g.][]{snellen2010orbital,birkby2018exoplanet}. This technique takes advantage of the fact that spectral lines at high resolution are Doppler-shifted due to the planet's orbital motion, allowing us to separate the planetary signal from the quasi-static stellar and telluric lines, thereby enabling the unambiguous detection of molecular and atomic species in the atmosphere of exoplanets.\\

With the help of cross-correlation techniques, we can compare numerous absorption or emission lines to model templates, which can significantly enhance the atmospheric signal and help constrain physical properties such as chemical abundances, temperature-pressure (\textrm{$T$-$P$}) profile \citep[e.g.][]{gibson2020detection, line2021solar, 2021AJ....162...73P}, and planetary rotation \citep[e.g.][]{2014Natur.509...63S,2015ApJ...814L..24L,brogi2016rotation,2023AJ....165..242G}. This has emerged as one of the most effective methods for characterising exoplanet atmospheres, particularly those of hot/ultra-hot Jupiters (UHJs; $T_\mathrm{eq}$\,$\gtrsim$\,$2{,}200$ K). Due to their close orbital periods to their parent stars, these tidally locked gas giants receive intense stellar irradiation. The day-side temperature thus increases to a level where their atmospheric and chemical structures are expected to differ significantly from those of cooler hot Jupiters \citep[e.g.][]{2018A&A...617A.110P}. Therefore, ultra-hot Jupiters serve as a treasure trove for numerous atomic, ionic, and molecular opacity sources (e.g. $\rm Fe$\hspace{0.3mm}\textsc{i}, $\rm Fe$\hspace{0.3mm}\textsc{ii}, $\rm Ca$\hspace{0.3mm}\textsc{i}, $\rm Ca$\hspace{0.3mm}\textsc{ii}, $\rm OH$, $\rm CO$, $\rm H_2O$, etc.), many of which have been detected using high-resolution transmission and emission spectroscopy \citep[e.g.][]{hoeijmakers2019spectral, gibson2020detection, nugroho2020detection, 2021MNRAS.506.3853M, 2023MNRAS.519.1030M, 2021ApJ...910L...9N, line2021solar, brogi2023roasting}.\\

Emission spectroscopy effectively enables a direct spectrum of the planet, which allows measurements of the temperature structure and chemical composition of planetary atmospheres. In addition, resolved phase-curve observations enable us to obtain spectra as the planet rotates, allowing us to effectively spatially-resolve the surface \citep[e.g.][]{2019ApJ...872....1R,2019MNRAS.488.1332L,2021MNRAS.501...78P}. The prevalence of thermal inversions, a rise in temperature with altitude, has served as a major driving force for much of the characterization work done on hot and ultra-hot Jupiters. Temperature inversions are seen in the Solar System on planets with substantial atmospheres \citep[e.g.][]{robinson2014common}. For example, ozone absorption in the stratosphere causes a temperature inversion on Earth, whereas hydrocarbons drive inversions in the stratospheres of Jupiter and Saturn. Heavily-irradiated gas giants (hot/ultra-hot Jupiters) were theorised to harbour temperature inversions due to the possible presence of strong ultra-violet and optical absorbers, such as $\mathrm{VO}$ and $\mathrm{TiO}$ (in gaseous form), which can exist at high temperatures in the upper atmosphere of these planets \citep{hubeny2003possible, fortney2008unified, molliere2015model, lothringer2018extremely, gandhi2019new}. Although the inversion agents are not yet well known \citep{sheppard2017evidence, arcangeli2018h}, many UHJs display temperature inversions, including WASP-33 b \citep{haynes2015spectroscopic, nugroho2017high, nugroho2020searching}, WASP-121 b \citep{evans2016detection}, and WASP-18 b \citep{sheppard2017evidence, brogi2023roasting}.\\

While the cross-correlation technique is extremely efficient at detecting atmospheric species, it is not possible to robustly compare the cross-correlation signals of various observations and model atmospheres to determine best-fitting models and obtain quantitative constraints on atmospheric parameters of interest. Furthermore, it does not provide a straightforward method for extracting the exoplanet's spectrum from the data, which can then be readily fitted with atmospheric models to carry out retrievals, as in the case of low-resolution observations. With recent developments in high-resolution Bayesian methods \citep[e.g.][]{brogi2019retrieving, gibson2020detection}, these observations enable detailed atmospheric retrievals, allowing us to recover important constraints on the atmosphere, such as the temperature-pressure profile and chemical/elemental abundances. This, in turn, can help place constraints on important quantities such as the carbon-to-oxygen ratio (henceforth $\rm{C/O}$) in planetary atmospheres, potentially providing new information on how planets form and evolve in the protoplanetary disk \citep{oberg2011effects, 2014ApJ...794L..12M, 2016ApJ...832...41M}.\\

The original CRyogenic high-resolution InfraRed Echelle Spectrograph \citep[CRIRES;][]{kaeufl2004crires} installed on the European Southern Observatory's (ESO) Very Large Telescope (VLT) was used to pioneer the field of high-resolution spectroscopy of exoplanetary atmospheres, obtaining the first detection of $\mathrm{CO}$ in transmission \citep{snellen2010orbital} and $ \mathrm{H_2O} $ in emission \citep{birkby2013detection} until its decommissioning in 2014. Recently, CRIRES has been upgraded to a cross-dispersed echelle spectrograph \citep{dorn2014crires+, follert2014crires+}. The improved spectrograph (now CRIRES+) offers a ten-fold improvement in spectral coverage within a wavelength range of \textrm{$0.95$--$5.3$} $ \upmu $m, and promises a leap in performance for high-resolution infrared spectroscopy alongside an increase in throughput of ${\approx}15\%.$\footnote{A detailed description of the instrument can be found in the CRIRES+ User Manual (P109.4): \url{https://www.eso.org/sci/facilities/paranal/instruments/crires/doc/ESO-254264_CRIRES_User_Manual_P109.4.pdf}} During the refurbishing of CRIRES+, other high-throughput spectrographs have become available. Although mounted on smaller telescopes, they offer a simultaneous snapshot of the entire NIR range (e.g. SPIRou, GIANO) or two out of the three bands (e.g. CARMENES, IGRINS). Therefore, it is important to evaluate the capabilities of CRIRES+ in comparison to these equally performing spectrographs.\\

Between September 15 and September 19 2021, CRIRES+ was used to obtain Science Verification (SV) observations with several programmes dedicated to transiting exoplanet observations, including WASP-20b (hot Saturn), HIP 65Ab (extreme hot Jupiter), and LTT 9779b (hot Neptune). Here, we present an analysis of high-resolution secondary eclipse observations of the ultra-hot Jupiter MASCARA-1b (also known as HD 201585b) from program 107.22TQ.001 (PI: Gibson). MASCARA-1b orbits a fast-rotating A8 star ($V$\,=\,$8.3$) with an orbital period of $2.15$ days \citep{talens2017mascara, hooton2022spi}. With an equilibrium temperature of $2{,}570$ K, MASCARA-1b is among the hottest and most highly irradiated exoplanets discovered to date. The spin axis of the host star, MASCARA-1, is misaligned with the planet’s orbit, with an obliquity of $ 69.5 \pm 0.3^{\circ} $, as is typical for hot Jupiters transiting early-type stars \citep{2010ApJ...718L.145W, 2010ApJ...719..602S, 2012ApJ...757...18A}. This ultra-hot Jupiter has been observed using high-resolution transmission spectroscopy with HARPS and ESPRESSO \citep{stangret2022high, casasayas2022transmission}, reporting non-detection of absorption features due to the presence of a strong Rossiter-McLaughlin (RM) effect, causing an overlap of any potential planetary signal with the Doppler shadow, in addition to its relatively small atmospheric scale height due to its high surface gravity. In contrast, MASCARA-1b's high day-side temperature makes it an excellent target for {\it emission} spectroscopy studies.\\

Emission spectroscopy of MASCARA-1b has already been reported by \citet[]{holmberg2022first} using CRIRES+ and \citet{2023arXiv230403328S} using PEPSI on the Large Binocular Telescope. We note that \citet[]{holmberg2022first} analyse the same CRIRES+ data we focus on here, reporting detections of CO and H$_2$O. \citet{2023arXiv230403328S} further detect the presence of Fe, Cr, and Ti in the atmosphere. In this paper, we use the CRIRES+ data and obtain higher-significance detections of $\rm CO$ and H$_2$O, plus detect the Fe signal in the K-band. This enables us to perform an atmospheric retrieval to constrain the metallicity and $\rm C/O$ ratio of the planet's atmosphere. In Section~\ref{sect2}, we describe the CRIRES+ observations and data reduction. We then describe our forward model atmosphere for emission and detail our methodology, including the cross-correlation technique and likelihood mapping approach, in Section~\ref{sect3}. In Section~\ref{sect4}, we present our detection results and the atmospheric retrievals. Finally, in Section~\ref{sect5}, we discuss our findings and present possible avenues to explore in future work before concluding the study and summarising our results in Section~\ref{sect6}.

\section{CRIRES+ Observations and Data Reduction}\label{sect2}
We observed the target MASCARA-1b during the second half of the night of 2021 September 16, using the upgraded CRIRES+ spectrograph \citep[$\rm R$\,$\sim$\,$100{,}000$; $\lambda$\,$\sim$\,$0.95$--$5.3$ $\upmu$m;][]{follert2014crires+} on UT3 (Unit Telescope 3) of the VLT, as part of program 107.22TQ.001 (PI: Gibson). We obtained $ 107 $ exposures covering the orbital phase of MASCARA-1b from $\phi$\,$\sim$\,0.33 to 0.42 (where $\phi$\,=\,0 corresponds to the central transit and $\phi$\,=\,0.5 corresponds to secondary eclipse). We observed in the K-band using the K2166 wavelength setting with (interrupted) coverage from $\lambda$\,$\sim$\,\textrm{$1921$--$2472$ $\rm nm$}. Table~\ref{tab:table1} provides a summary of the observations.\\

\begin{figure}
    \includegraphics[width=\columnwidth]{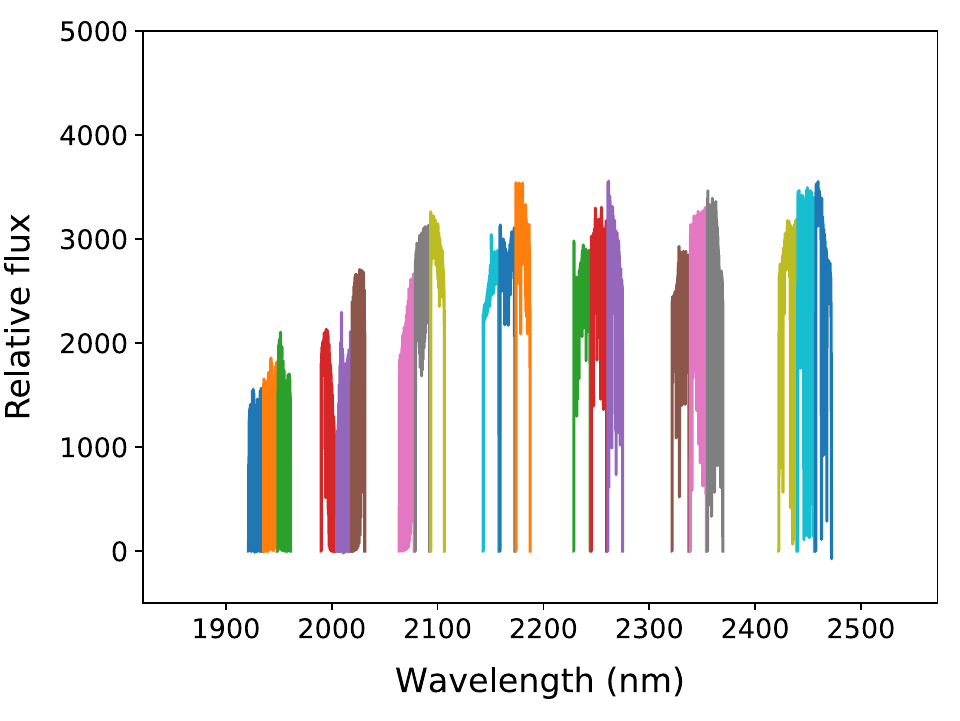}
    \caption{An example of the reduced CRIRES+ data for a single spectrum.}
    \vspace{3pt}
    \label{fig:data}
\end{figure}

The data were analysed and reduced using the ESO {\sc cr2res} data pipeline (version 1.1.4) and executed via the Recipe Execution Tool {\sc EsoRex}\footnote{Documentation available at the ESO website \url{http://www.eso.org/sci/software/pipelines/}}, which performed standard calibrations and extracted the time-series spectra for each spectral order. The pipeline performed dark subtraction, flat-field correction, and wavelength calibration (for more information, see the CRIRES+ Pipeline Manual v1.2.3). In summary, we combined each set of raw dark frames into master dark frames (which also produced a bad pixel map, BPM). Next, we compute master flat frames and perform trace detection. Following this, we perform wavelength calibration of the extracted spectra using the Fabry-Pérot Etalon (FPET) frame. Finally, we calibrate and extract the science spectra (1D spectra as a function of order).
\begin{table}
	\centering
	\caption{An overview of MASCARA-1b observations during the Science Verification run of CRIRES+ ($ \mathrm{N_{obs}} $ is the total number of observed spectra, the exposure time is expressed as NDIT\,$\times$\,DIT, where $ \mathrm{DIT, NDIT} $ corresponds to the detector integration time and number of detector integrations, respectively).}
	\label{tab:table1}
	\begin{tabular}{lc} 
		\hline
		Target & MASCARA-1b\\
        Programme ID & 107.22TQ.001\\
        PI & Gibson\\
        Night & 2021-09-16\\
        Phase, $ \phi $ & \textrm{$0.32$--$0.42$}\\
        $ \mathrm{N_{obs}} $ & 107\\
        Exp. Time & 5\,$\times$\,30 $\rm s$\\
        Obs. Mode & Staring\\
        Slit & $ 0.2'' $\\
        AO loop & Closed\\
        Wavelength Setting & $ \mathrm{K2166} $ (\textrm{$1921$--$2472$} $\rm nm$)\\
		\hline
	\end{tabular}
\end{table}
In total, we obtain 7 echelle orders for the K2166 wavelength setting. In addition, with CRIRES+, each spectral order is observed across 3 separate detectors (CHIP1, CHIP2, and CHIP3). We, therefore, treat each separate chip independently and hereafter refer to each of these as `orders'. We finally produce a 3D data cube (order\,$\times$\,time/phase\,$ \times$\,wavelength) with 21 spectral orders ranging from $\lambda$\,$\approx$\,1920 to 2470 $\mathrm{nm}$. The average signal-to-noise ratio (S/N) at the centre of each order is $\approx$100. For each exposure, we calculated the barycentric velocity correction using the online tool from \citet{2014PASP..126..838W} and calculated the orbital phase using the transit epoch taken from \citet{talens2017mascara}. An example of the reduced CRIRES+ data for a single spectrum is shown in Fig.~\ref{fig:data}.

\subsection{Order pre-processing}\label{sec:2.1}
We perform a series of pre-processing steps and begin by removing outliers from each order of the extracted spectra by subtracting a model for the data constructed from the outer product of the spectral median (i.e. median over time) and median light curve (i.e. median over wavelength), divided by the overall mean to normalise. Before adding the cleaned array to the model, each residual spectrum was fitted with a $\mathrm{10^{th}}$-order polynomial, and any outliers $>$4\,$\sigma$ were replaced by their corresponding polynomial value. While this is an arbitrary threshold, this procedure replaced an average of approximately 0.03\% of pixels per order, and, therefore, the exact choice of threshold or polynomial order has negligible impact.\\

Following the procedure outlined in \citet{gibson2020detection}, we extracted estimates of the noise for each order by assuming a Poisson-dominated noise term, with the standard deviation $ \sigma_i = \sqrt{aF_i + b} $ where $ F_i $ is the measured flux for a given time and wavelength, and the coefficients $a$ and $b$ denote a gain and read noise, respectively. To extract the noise in each order, we subtract a $\mathrm{2^{nd}}$-order Principle Component Analysis (PCA) model for each order of the cleaned array. This gives us a residual array, $ R_\mathrm{i}$, and the values for $ a $ and $ b $ were found by fitting our noise model (i.e. $ \sigma_i $) to each order by optimising the Gaussian log-likelihood of the form:
\begin{equation*}
    \ln \mathcal{L}(a,b) =  -0.5\sum_{i}\left(\frac{R_{i}}{\sigma_i}\right)^2 - \sum_{i} \ln \sigma_i
\end{equation*}
To determine the uncertainties, we use the best-fitting values for $a$ and $b$ and then fit this estimate with a $\mathrm{2^{nd}}$-order PCA model to remove any bias in the noise estimation (see \citealt {gibson2020detection} for further details), making this model our final estimate of the uncertainties. We note that while this approach is better suited for prior studies focused on optical data \citep[e.g.][]{gibson2020detection}, it might not capture the final noise amplitude for poorly-corrected telluric, stellar, or systematic effects in the NIR data. Therefore, as an alternative to the optimisation presented above, we also estimate the uncertainties of each order post-\textsc{SysRem} by taking the outer product of the standard deviation of each wavelength and spectrum, normalised by the overall mean. We then use the new uncertainties to re-run \textsc{SysRem} and find that the choice of error estimation method did not have a discernible impact on our detections or retrieved values. Additionally, we also note that errors set via the $\rm (a,b)$ optimisation resulted in slightly higher detection significance ($\approx$17$\sigma$) compared to the latter ($\approx$16$\sigma$).\\

Following the procedure outlined in \citet{gibson2019revisiting}, we apply a blaze correction to the resultant spectral orders by first dividing each spectrum by the spectral median in each order, then smoothing the resulting spectral residuals with a median filter with a width of 501 pixels and a Gaussian filter with a standard deviation of 100 pixels, creating a smoothed map of the blaze distortion per order. For accurate retrievals, it is essential to ensure that the blaze correction only removes gradual and consistent changes in the blaze caused by systematic factors without affecting the underlying exoplanet signal. Therefore, similarly to \citet{gibson2022relative}, we used a significantly wider kernel width and standard deviation for the median and Gaussian filters. Finally, each of the original spectra (and corresponding uncertainties) was divided by their respective blaze correction. This procedure does not remove the blaze function but places every spectrum (for each order) on a `common' blaze. To ensure this process did not significantly distort the underlying exoplanet signal, we performed injection tests with an atmospheric model containing $\mathrm{{}^{12}CO}$, $\mathrm{H_2O}$ and $\mathrm{Fe}$. The model used the negative value of expected $K_{\rm p}$ to ensure that the injected signal is well separated from the real exoplanet signal. We performed a retrieval analysis on the injected data (see Section~\ref{sec:3.6}) and found the retrieved model parameters, as well as the retrieved abundances and \textrm{$T$-$P$} profiles, to be in agreement with the injected values, therefore confirming that our blaze correction does not bias the results. These results are outlined in Figure.~\ref{fig:injection_test}.

\subsection{Removal of stellar and telluric features}\label{sec:2.2}
\begin{figure}
    \includegraphics[width=1.07\columnwidth]{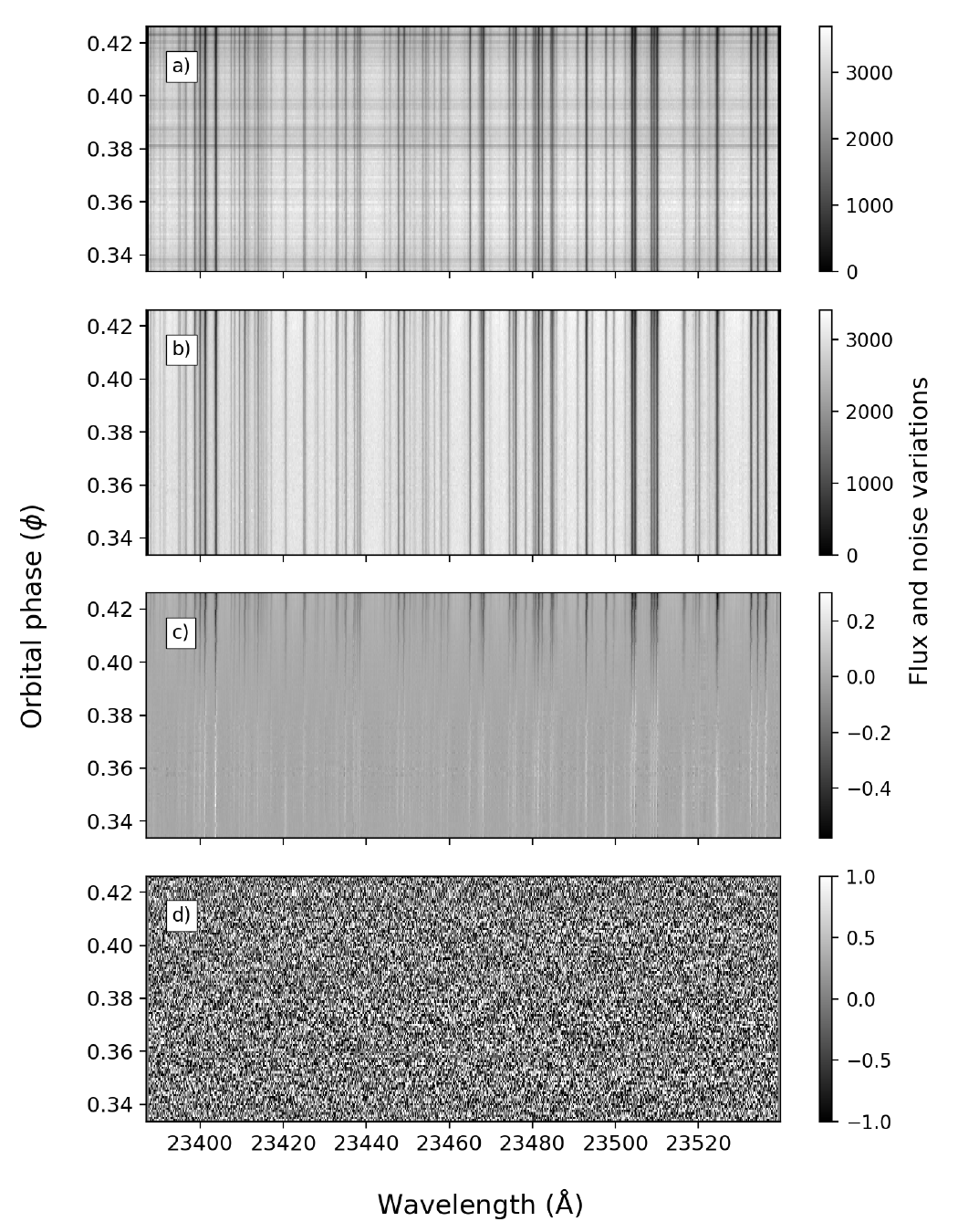}
    \caption{The data pre-processing steps applied to a single order, a) cleaning and outlier removal, b) after wavelength-shift and blaze correction, c) the global \textsc{SysRem} model, and d) after division by the median spectrum and subtraction of the \textsc{SysRem} model.}
    \vspace{3pt}
    \label{fig:sysrem}
\end{figure}

To search for the buried and Doppler-shifted exoplanetary signal, all trends in the data which are (quasi-)static in time must be removed. Therefore, to remove the stellar and telluric features we use the \textsc{SysRem} algorithm \citep{tamuz2005correcting}, which was first adapted to high-resolution spectroscopy by \citet{birkby2013detection} and has since then been successfully applied to both transmission and emission spectra \citep[e.g.][]{birkby2017discovery, nugroho2017high, nugroho2020searching, nugroho2020detection, gibson2020detection, gibson2022relative, 2023MNRAS.519.1030M}. We follow the procedure outlined in \citet{gibson2022relative}, and first normalise the data by dividing each order through by the median spectrum before applying \textsc{SysRem}. For each \textsc{SysRem} iteration (and order), the time $\times$ wavelength data array is decomposed into two column vectors, $ \boldsymbol{u} $ and $ \boldsymbol{w} $, where the model array for each pass $ i $ is determined by their outer product $\boldsymbol{u_i}\boldsymbol{w_i}^{\rm T}$. After one iteration, the resultant model is subtracted from the input data to produce the processed data, and the procedure is repeated for the subsequent iteration. Thus, the \textsc{SysRem} model for a single order, $ D $, after $ N $ iterations is:
\begin{equation}
    D =  \sum_{i=1}^{N}\boldsymbol{u_i}\boldsymbol{w_i}^{\rm T} = {\mathbfss{U}\mathbfss{W}^{\rm T}}
    \label{eq:1}
\end{equation}
where $\mathbfss{U}$ and $\mathbfss{W}$ are matrices containing column vectors $\boldsymbol{u_i}$ and $\boldsymbol{w_i}$. This final \textsc{SysRem} model is then subtracted from the normalised data, and the matrix $\mathbfss{U}$ is stored for processing the forward model (see Sect.~\ref{sec:3.3}). Unless otherwise stated, we apply 15 passes of \textsc{SysRem}. While that is a somewhat arbitrary selection, we ran further tests by repeating our retrievals with $N$\,=\,5,\,10,\,15 and 20 \textsc{SysRem} iterations. We find that using $N$\,=\,5,\,10 and 15 gives us consistent results, whereas $N$\,=\,20 filtered out some of our exoplanet signal (see Section~\ref{sec:4.2}). Lastly, the uncertainties for each order determined earlier are divided through by the median spectrum to account for the pre-processing (subtraction of the \textsc{SysRem} model does not modify the uncertainties). A step-by-step overview of the pre-processing steps we apply is shown in Fig.~\ref{fig:sysrem} for a single order.

\section{Methods}\label{sect3}
\subsection{Model emission spectra}\label{sec:3.1}
The high-resolution cross-correlation technique requires spectral templates that are used to search for atomic and molecular species in the atmosphere of exoplanets. Both cross-correlation and likelihood approach require an accurate template to optimise the detection significance of species and make reliable quantitative constraints on atmospheric parameters. The likelihood approach further demands that the forward model is extremely fast so that many forward models can be generated at speed to fully explore the posterior distribution.\\

For our analysis, we, therefore, used the atmospheric code \textsc{irradiator} \citep{gibson2022relative}. This was initially developed to compute transmission spectra at high-resolution and was accelerated by re-writing much of the radiative transfer calculations as matrix-vector or matrix-matrix products where possible and using single-precision floating-point calculations, which result in negligible loss of accuracy when compared to the discretization of layers and interpolation of cross-sections. For this work, we extended the \textsc{irradiator} code to compute emission spectra and also compute chemical equilibrium models using the chemical code \textsc{FastChem} \citep{stock2018fastchem}\footnote{\textsc{FastChem} is available at \url{https://github.com/exoclime/FastChem}}. Here we briefly describe some of the updates to \textsc{irradiator} and refer the reader to \citet{gibson2022relative} for further details. We first define a set of atmospheric layers covering a range of pressures (uniform in log space) and define/compute the \textrm{$T$-$P$} profile at each layer (see Sect.~\ref{sec:3.2}). For each species of interest, we then specify the volume mixing ratio (VMR), $\chi_{\rm species}$, at each layer. These are either given as free parameters for each species (in which case the species are assumed to be well-mixed) or computed using \textsc{FastChem} for each layer of the atmosphere (taking into account the temperature and pressure) and using an additional free parameter to specify the metallicity (relative to solar) and $\rm C/O$ ratio (where we adjust both $\rm C$ and $\rm O$ abundances to have a desired ratio but preserve their sum). Once the \textrm{$T$-$P$} profile and abundances have been set, we then compute the vertical structure of the atmosphere by assuming hydrostatic equilibrium. We then compute the opacity at each layer of the atmosphere using a set of pre-computed opacity grids over temperature and pressure, which we linearly interpolate for each layer and sum the contributions from each species. We then integrate these vertically through the atmosphere to compute the transmission from each layer out to space. We finally use standard radiative transfer equations \citep[e.g.][]{2010ppc..book.....P,molliere2019petitradtrans} to compute the emission for every point on the wavelength grid by integrating through the atmosphere and assuming black-body emission from the deepest layer. Here we compute spectra across a wavelength range of \textrm{$18000$--$28500$ \AA} at a constant resolution of $\mathrm{R}$\,=\,$200{,}000$. We compute the emission spectrum (in units of spectral irradiance) at a range of angles through the atmosphere before integrating to get the emergent flux. Similarly to \citet{molliere2019petitradtrans}, we use 3-point Gaussian quadrature.\\

Our previous application of \textsc{irradiator} only implemented a simple cloud deck and Rayleigh scattering as sources of continuum opacity. Here we also add support for the bound-free and free-free absorption from $\mathrm{H^{-}}$ and collisionally-induced absorption (CIA) of H$_2$-H$_2$ and H$_2$-He which are potentially important for this temperature regime and wavelength range. We use the cross-sections from \citet{2005oasp.book.....G} and \citet{doi:10.1021/jp109441f, 10.1063/1.3676405}. The VMRs of the relevant species (i.e. H, H$^-$, e$^-$, He, H$_2$) are either specified in advance (i.e. free parameters) or computed using \textsc{FastChem}. Finally, the emission spectrum for the planet in relation to the star (planet-to-star flux ratio) is computed as:
\begin{equation*}
    \frac{F_\mathrm{p}(\lambda)}{F_\mathrm{*}(\lambda)} = \frac{F_\mathrm{model}}{\pi B(\lambda, T_{\mathrm{eff}})} \left( \frac{R_\mathrm{p}}{R_\mathrm{*}} \right)^2
\end{equation*}
where $ F_\mathrm{model} $ is the model flux in $ \mathrm{Wm^{-2}\hspace{0.5mm}m^{-1}} $, $ B $ denotes the Planck function at the effective temperature ($ T_{\mathrm{eff}} $) of the star\footnote{The factor of $\pi$ accounts for the conversion of spectral irradiance to flux assuming a Lambertian surface}, and $ R_\mathrm{p} $ and $ R_\mathrm{*} $ denote the planetary and stellar radii, respectively. For the opacities, we focus on molecular and atomic species that are known to dominate the atmosphere of ultra-hot Jupiters at near-infrared wavelengths \citep[e.g.][]{snellen2010orbital, birkby2013detection, de2013detection, nugroho2020detection, line2021solar}. For the remainder of this work, we use the pre-computed opacity grids provided by {\sc petitRADTRANS}\footnote{\url{https://petitradtrans.readthedocs.io/en/latest/}} \citep{molliere2019petitradtrans, molliere2020retrieving}. We also include a parameterized cloud deck pressure ($ P_\mathrm{cloud} $) where we assume the atmosphere has infinite opacity below.\\

\begin{figure}
	\includegraphics[width=\columnwidth]{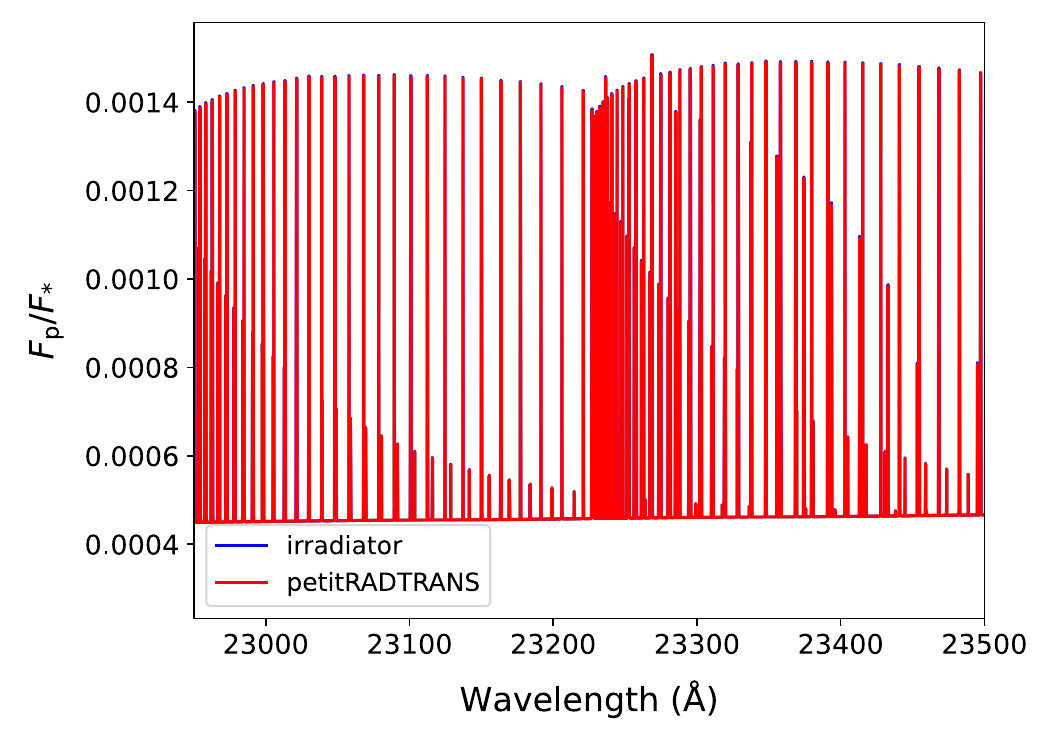}
    \caption{An example of our forward model atmosphere ($ \mathrm{{}^{12}CO} $) in units of planet-to-star flux ratio. The blue line is our model, and the red is the corresponding {\sc petitRADTRANS} model. These are barely distinguishable, and similarly to the transmission case, they are likely due to differences in numerical approaches to computing the vertical structure and radiative transfer.}
    \label{fig:prtvsirr}
\end{figure}

Similarly to \citet{gibson2022relative}, we bench-marked our models against {\sc petitRADTRANS}. An example of our forward model atmosphere for $ \rm{{}^{12}CO} $ (VMR\,=\,8\,$\times$\,$10^{-4}$) with a cloud deck at 1 bar is shown in Fig.~\ref{fig:prtvsirr}. We used a model atmosphere with 100 layers and an inverted $T$-$P$ profile (middle panel of Fig.~\ref{fig:bfmodel_GuiTP}). The equivalent model computed with {\sc petitRADTRANS} is over-plotted, showing that models are consistent and most likely the result of different numerical approaches, e.g. solving hydrostatic equilibrium in log-space.

\subsection{Temperature-Pressure profile}\label{sec:3.2}
Following the procedure described above, we define a series of atmospheric layers that span a range of pressures (evenly spaced in logarithmic units) and compute the temperature-pressure (\textrm{$T$-$P$}) profile using the parametric model from \citet[][Eqn. 29]{guillot2010radiative} that has been widely used by many authors \citep[e.g.][]{brogi2019retrieving, molliere2019petitradtrans, nugroho2020detection, gibson2022relative, 2023MNRAS.519.1030M}. This \textrm{$T$-$P$} profile allows for inverted or non-inverted atmospheres and is a relatively simple parameterization with four terms: the irradiation temperature $ T_\mathrm{irr} $, the mean infrared opacity $\kappa_{\rm IR}$, the ratio of visible-to-infrared opacity $\gamma$, and the internal temperature $ T_{\rm int} $. We adopt a thermally inverted \textrm{$T$-$P$} profile \citep{guillot2010radiative}, assuming the internal temperature ($ T_\mathrm{int} $) of 100 K, $ T_{\rm irr} $ of 2600 K, and the mean infrared opacity ($ \kappa_\mathrm{IR} $) of 0.01 $\rm m^2$ $\rm kg^{-1}$. While physically motivated, the \citet[][]{guillot2010radiative} profile can be restrictive in setting the gradient of the \textrm{$T$-$P$} profile when compared to more empirical methods. Therefore, we also implement the approach of \citet{madhusudhan2009temperature} where we divide the atmosphere into 3 layers  as follows:
\begin{equation}
  T(P) =
    \begin{cases}
      \frac{1}{\alpha_1} = \ln^2 \left( \frac{P}{P_0} \right) + T_0 & P_0 \le P \le P_1\\
      \frac{1}{\alpha_2} = \ln^2 \left( \frac{P}{P_2} \right) + T_2 & P_1 \le P \le P_3\\
      T = T_3 & P \ge P_3
    \end{cases}       
\end{equation}
where temperatures $ T_2 $ and $ T_3 $ are determined via continuity at $ P_1 $ and $ P_3 $, respectively \citep{line2021solar}. Finally, the profile is smoothed according to the number of layers using a 1D uniform filter (see \citet{madhusudhan2009temperature} for a detailed description). This profile takes six free parameters as inputs: $T_0$, $\alpha_1$, $\alpha_2$, $P_1$, $P_2$, and $P_3$, corresponding to temperature at the top of the atmosphere, parameters governing the change of temperature with pressure in each layer, and pressures of layers 1, 2, and 3, respectively. The implementation of this parametric model allows more flexibility in setting the temperature gradient, which here is governed by parameters $ \alpha_1 $ and $ \alpha_2 $ (see, e.g. Fig.~\ref{fig:chemmodel_mstp}).\\
\begin{figure*}
  \includegraphics[width=1.95\columnwidth]{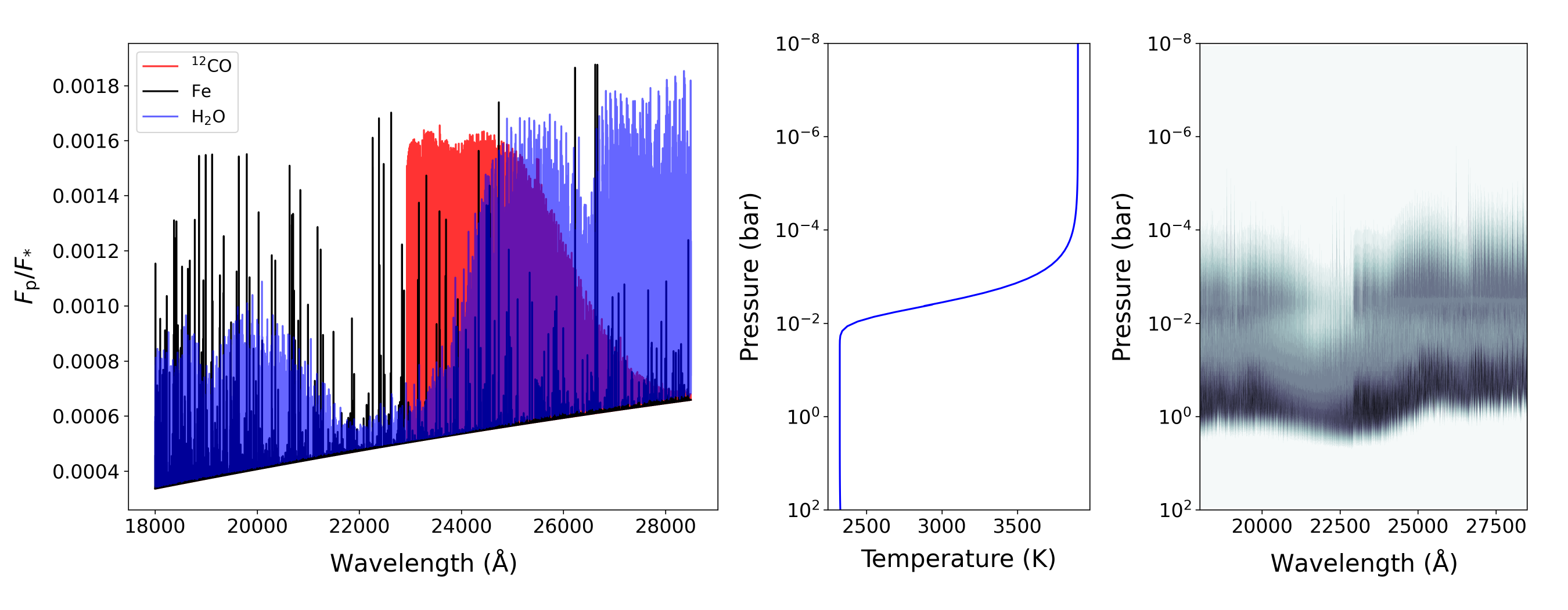}
  \caption{Model emission spectrum of MASCARA-1b with contributions from $ \mathrm{CO} $, $ \mathrm{H_2O} $ and $ \mathrm{Fe} $ shown individually ($\log_{{10}} \chi_{\rm species} = -4$; set using the maximum abundance computed at any pressure level from a chemical model). \textit{Middle panel:} parametric $T$-$P$ profile from \citet{guillot2010radiative}. \textit{Right panel:} combined emission contribution function (pressures greater than $ {\approx}1 $ bar cannot be probed).}\label{fig:bfmodel_GuiTP}
\end{figure*}

In summary, our 1D atmosphere can be specified using two different parametric \textrm{$T$-$P$} profiles (one physically motivated and one empirical) and two different chemical regimes -- either assuming the species of interest are well-mixed (constant VMRs with altitude) or by using {\sc FastChem} to compute the chemical profiles after specifying a metallicity and $\rm C/O$ ratio. We apply various combinations of these parameterizations to both our cross-correlation and retrieval analyses, which we discuss in more detail in Section~\ref{sec:4.2}. The model spectrum and corresponding \textrm{$T$-$P$} profile are shown in Fig.~\ref{fig:bfmodel_GuiTP}.

\subsection{Model filtering}\label{sec:3.3}
The data is subjected to a number of pre-processing steps, as described in Sections~\ref{sec:2.1} and~\ref{sec:2.2}, to remove the stellar and telluric features and various instrumental artefacts (such as blaze corrections, bad pixels, etc). Most importantly, this will also alter the underlying exoplanet signal. Therefore, for our forward model to accurately represent the data, it is necessary to apply the same pre-processing methods as the data to the model.\\

We implement the novel model-filtering technique introduced by \citet{gibson2022relative}, which makes use of the output matrices \textbfss{U} and \textbfss{W} (Section~\ref{sec:2.2}), containing the column and row vectors $\boldsymbol{u}$ and $\boldsymbol{w}$ for each \textsc{SysRem} iteration. To account for the fact that we are unsure of the precise broadening caused by, for example, winds or rotation, we first broaden our model emission spectra (Section~\ref{sec:3.1}) via convolution with a Gaussian kernel with a standard deviation, $ W_\mathrm{conv} $, enabling constant velocity broadening (as the wavelength of the model is sampled at constant resolution). The convolved model is then linearly interpolated to the wavelength grid of our data (order\,$\times$\, wavelength) and then Doppler-shifted to a planetary velocity for each order:
\begin{equation}
    v_\mathrm{p} = K_\mathrm{p} \sin(2\pi\phi) + v_\mathrm{sys} + v_\mathrm{bary} \label{eq:vp}
\end{equation}
where $ K_\mathrm{p} $ is the radial velocity semi-amplitude of the planet's orbit, $ v_\mathrm{sys} $ is the systemic velocity offset, $ v_\mathrm{bary} $ is the barycentric velocity, and $ \phi $ is the orbital phase. This results in a 3D shifted forward model (phase/time\,$\times$\,order\,$\times $\,wavelength). The matrix multiplication in Eqn.~\ref{eq:1} can be considered as a linear basis model, where \textbfss{U} contains the $ N $ basis vectors for each \textsc{SysRem} iteration, and \textbfss{W} contains the corresponding weights. We fit the basis models \textbfss{U} to the 2D model atmosphere for each order by fixing the matrix \textbfss{U} and computing the weights using linear least squares. The best-fitting model is then simply the outer product of the best-fitting weights and the fixed basis vector $ \mathrm{\mathbfss{U}} $. Finally, to account for data uncertainties in the fit that were initially accounted for when computing $ \mathrm{\mathbfss{U}} $, we take the mean of the uncertainties for each order over wavelength $ \hat{\mathbf{\sigma}} $. We refer the reader to \citet{gibson2022relative} for a detailed description of the model filtering technique.

\subsection{Cross-correlation}\label{sec:3.4}
With our data free from stellar and telluric contamination and our shifted model filtered to imitate the impact of our pre-processing steps, we can now perform the traditional cross-correlation analysis \citep[e.g.][]{snellen2010orbital, gibson2020detection, 2021MNRAS.506.3853M} to extract the buried planetary signal. To generate a cross-correlation function (CCF), we multiply the data and the shifted model and sum over both wavelength and spectral order. This takes the following mathematical form:
\begin{equation}
    \mathrm{CCF} \hspace{0.2mm} (v_\mathrm{sys}) = \sum_{i=1}^{N}\frac{f_im_i(v_\mathrm{sys})}{\sigma_i^2}\label{eq:6}
\end{equation}
where the product is weighted on the variance of the data ($ \sigma_i^2 $ in Eqn.~\ref{eq:6}) while taking noise into account. The above equation produces cross-correlation values for each combination of orbital phase and systemic velocities (Eqn.~\ref{eq:vp}), referred to as a phase-velocity map or simply a ``cross-correlation'' map. The change in the radial velocity of the planet results in a Keplerian feature (a planetary trail) that is easily discernible in the cross-correlation map for strong planetary signals (e.g. see Fig.~\ref{fig:combinedmaps}), thereby allowing us to confidently confirm the presence of species in the atmospheres of exoplanets.\\

For weaker signals where the planetary trail is not visible, it is essential to integrate the cross-correlation map over a range of predicted planetary velocities $v_\mathrm{p}$. Typically, a range of radial velocity semi-amplitude values, $K_\mathrm{p}$, close to the predicted value (from radial-velocity measurements) are selected. Following this, for a given $K_\mathrm{p}$, the cross-correlation function for each orbital phase is shifted to a new planetary velocity (according to Eqn.~\ref{eq:vp}) and integrated over time to produce a \textrm{$K_{\rm p}$-$v_{\rm sys}$} map. By integrating cross-correlation functions across a range of planetary radial velocities, the source of the signal in velocity space can be pinpointed and compared with expected values, leading to the detection of a specific species in a planetary atmosphere. To place constraints on the signal amplitude, referred to as the detection significance, we subtract the map by the mean (in regions away from the peak) before dividing through by the standard deviation \citep{brogi2012signature, brogi2018exoplanet}. However, due to the arbitrary selection of this region, the resulting detection significance is not exact, implying that the same models and observations can lead to varied values for the detection significance.

\subsection{Likelihood mapping}\label{sec:3.5}
\begin{figure}
\begin{subfigure}{\columnwidth}
   \includegraphics[width=\columnwidth]{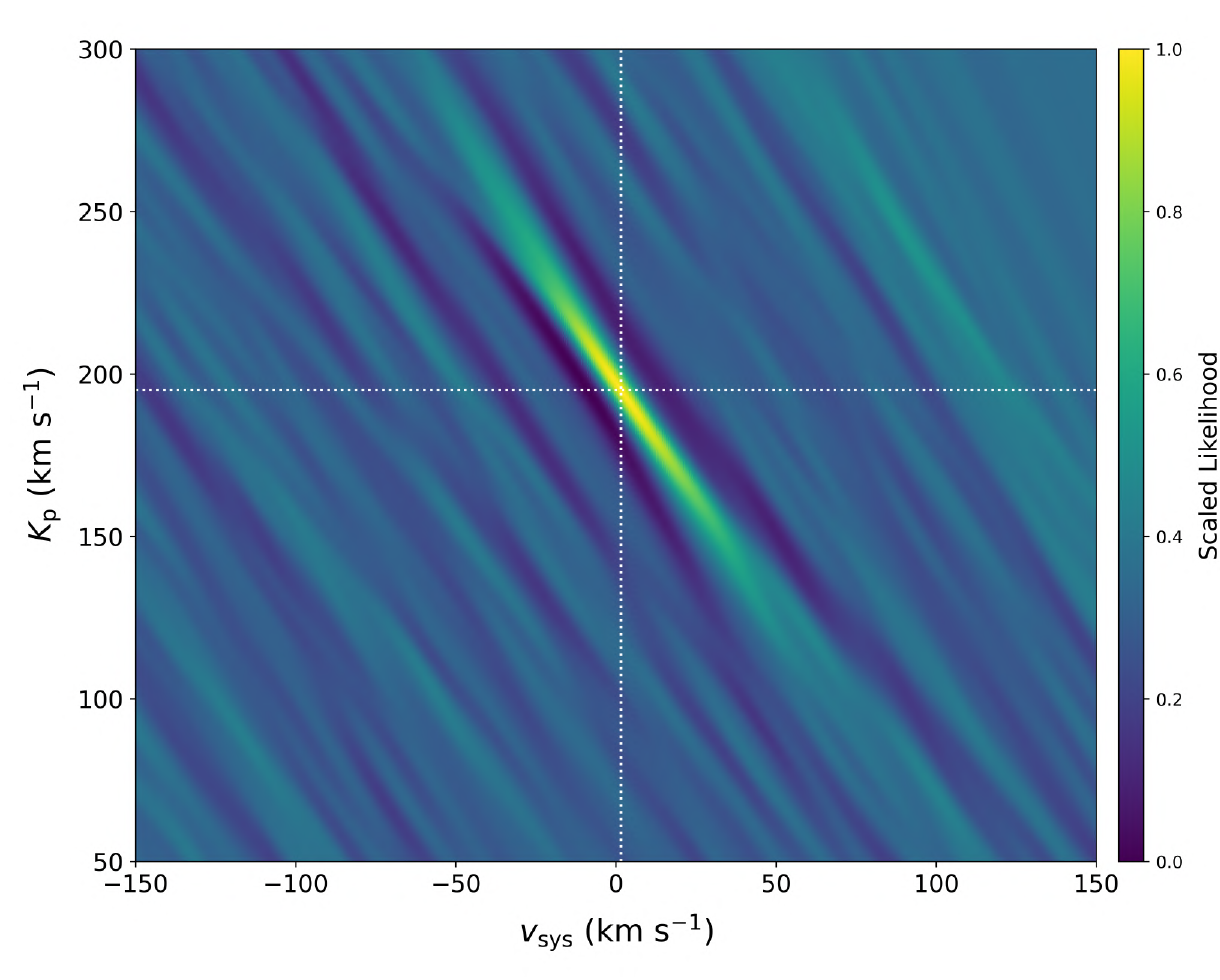}
\end{subfigure}

\begin{subfigure}{\columnwidth}
   \includegraphics[width=\columnwidth]{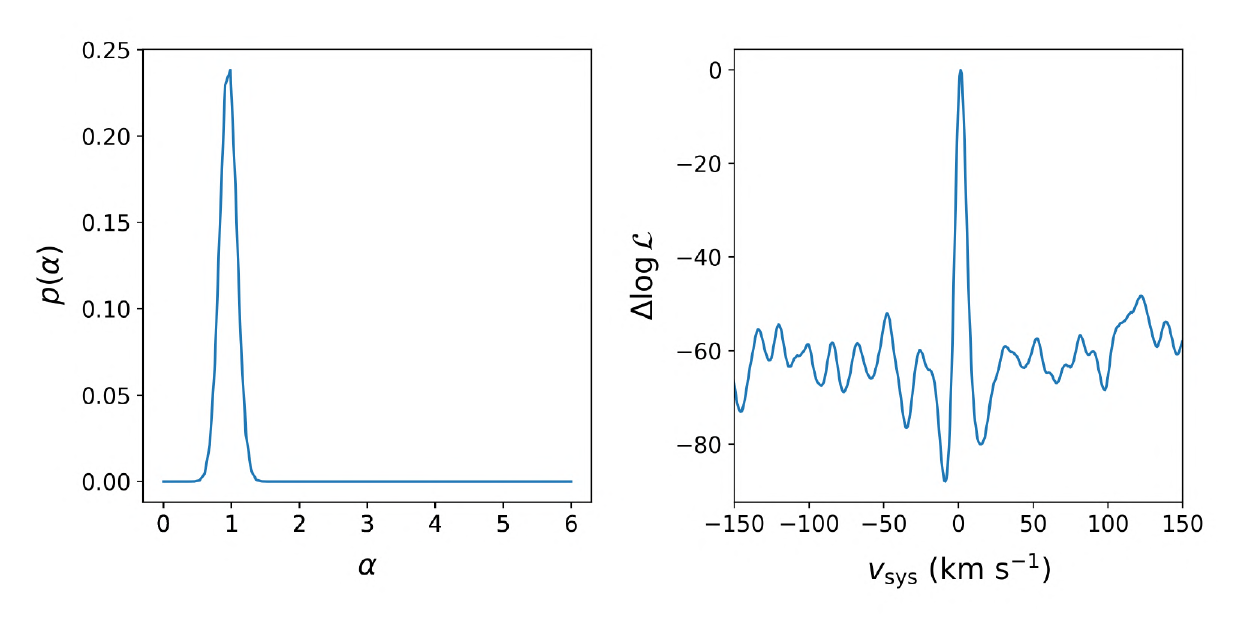}
\end{subfigure}
\caption{The combined likelihood map and conditional likelihood distribution of $ \alpha $ for $ \mathrm{CO} $, $ \mathrm{H_2O} $, and $ \mathrm{Fe} $. The top panel is the maximum likelihood map, normalised for visualisation. The white dotted lines mark the positions of maximum $ K_\mathrm{p} $ and $ v_\mathrm{sys} $. The bottom panel shows the conditional distribution of $ \alpha $, and a slice through the maximum $ K_\mathrm{p} $.}\vspace{23pt}\label{fig:bestfit_likelihood_maps}
\end{figure}

Despite being effective at distinguishing atomic and molecular properties in planetary atmospheres, the cross-correlation method does not allow direct comparisons between various model atmospheres. Therefore, \citet{brogi2019retrieving} first introduced a method to ``map'' cross-correlation values of a given atmospheric model to a likelihood value. \citet{gibson2020detection} developed an alternate but similar approach by employing a full Gaussian likelihood function that accounts for both wavelength- and time-dependent uncertainties. We will briefly outline this method here; for a detailed description, see \citet{gibson2020detection}. Starting with a full Gaussian likelihood function, with uncertainties that vary in time and wavelength:
\begin{equation}
    \mathcal{L} = \prod_{i=1}^{N} \frac{1}{\sqrt{2\pi (\beta\sigma_i)^2}}\exp{\left(-\frac{1}{2}\frac{(f_i - \alpha m_i(\theta))^2}{\beta\sigma_i^2}\right)} \label{eq:7}
\end{equation}
where $ \alpha $ and $ \beta $ denote the model scale factor and noise scale factor, respectively. A vector of model parameters is represented by $ \theta $, and $ i $ is indexed over wavelength, spectral order, and time. Dropping the reference to $ \theta $, the natural logarithm of the likelihood, or log-likelihood, is then computed as follows:
\begin{equation}
    \ln\mathcal{L} = -\frac{N}{2}\ln 2\pi - \sum_{i=1}^{N}\ln \sigma_i - N\ln \beta - \frac{1}{2}\chi^2 \label{eq:8}
\end{equation}
where,
\begin{equation}
    \chi^2 = \sum_{i=1}^{N}\frac{(f_i-\alpha m_i)^2}{(\beta \sigma_i)^2} \label{eq:9}
\end{equation}
The first two terms in Eqn.~\ref{eq:8} are constant for a given data set and thus can be dropped, giving:
\begin{equation}
    \ln\mathcal{L} = -N\ln \beta - \frac{1}{2}\chi^2 \label{eq:10}
\end{equation}
Expanding Eqn.~\ref{eq:9} gives:
\begin{equation}
    \chi^2 = \frac{1}{\beta}\left(\sum_{i=1}^{N} \frac{f_i^2}{\sigma_i^2} + \alpha^2 \sum_{i=1}^{N}\frac{m_i^2}{\sigma_i^2} - 2\alpha \sum_{i=1}^{N}\frac{f_im_i}{\sigma_i^2} \right) \label{eq:11}
\end{equation}
The final summation in Eqn.~\ref{eq:11} is equivalent to the CCF (Eqn.~\ref{eq:6}) summed over time, outline in Section~\ref{sec:3.3}, such that:
\begin{equation}
    \chi^2 = \frac{1}{\beta}\left(\sum_{i=1}^{N} \frac{f_i^2}{\sigma_i^2} + \alpha^2 \sum_{i=1}^{N}\frac{m_i^2}{\sigma_i^2} - 2\alpha \hspace{0.5mm} \mathrm{CCF} \right) \label{eq:12}
\end{equation}
Equations~\ref{eq:10} and~\ref{eq:12} allow the log-likelihood to be computed directly from the CCF, enabling direct model comparison and allowing the cross-correlation method to be folded into a Bayesian framework. By performing atmospheric retrievals from high-resolution observations, we can place constraints on the abundances of species, the atmospheric temperature structure, $\rm C/O$ ratios, etc. The likelihood map and a conditional likelihood distribution of $ \alpha$\footnote{We note that alpha is fixed to 1 in our retrievals but can be computed within the likelihood mapping, so we plot the marginalised posterior of alpha as a check.} are shown in Fig.~\ref{fig:bestfit_likelihood_maps}.

\subsection{Atmospheric Retrieval}\label{sec:3.6}
To perform an atmospheric retrieval, we must compute a forward model for a set of model parameters denoted by $\theta$. These parameters include \{$\alpha$, $K_{\rm p}$, $v_{\rm sys}$, $W_{\rm conv}$\}, where $ W_{\rm conv} $ is the width of the Gaussian broadening kernel in pixels. We also need to consider the input parameters of \textsc{irradiator}: either \{$\kappa_{\rm IR}$, $\gamma$, $T_{\rm irr}$, $P_{\rm cloud}$, $\chi_{\rm species}$\} or \{$T_0$, $\alpha_1$, $\alpha_2$, $P_1$, $P_2$, $P_3$, $P_{\rm cloud}$, $\chi_{\rm species}$\} for the two different parametric $T$-$P$ profiles, respectively \citep[][see Section~\ref{sec:3.1}]{guillot2010radiative, madhusudhan2009temperature}.
\begin{figure*}
\begin{minipage}[t]{0.333\textwidth}
  \includegraphics[width=\linewidth]{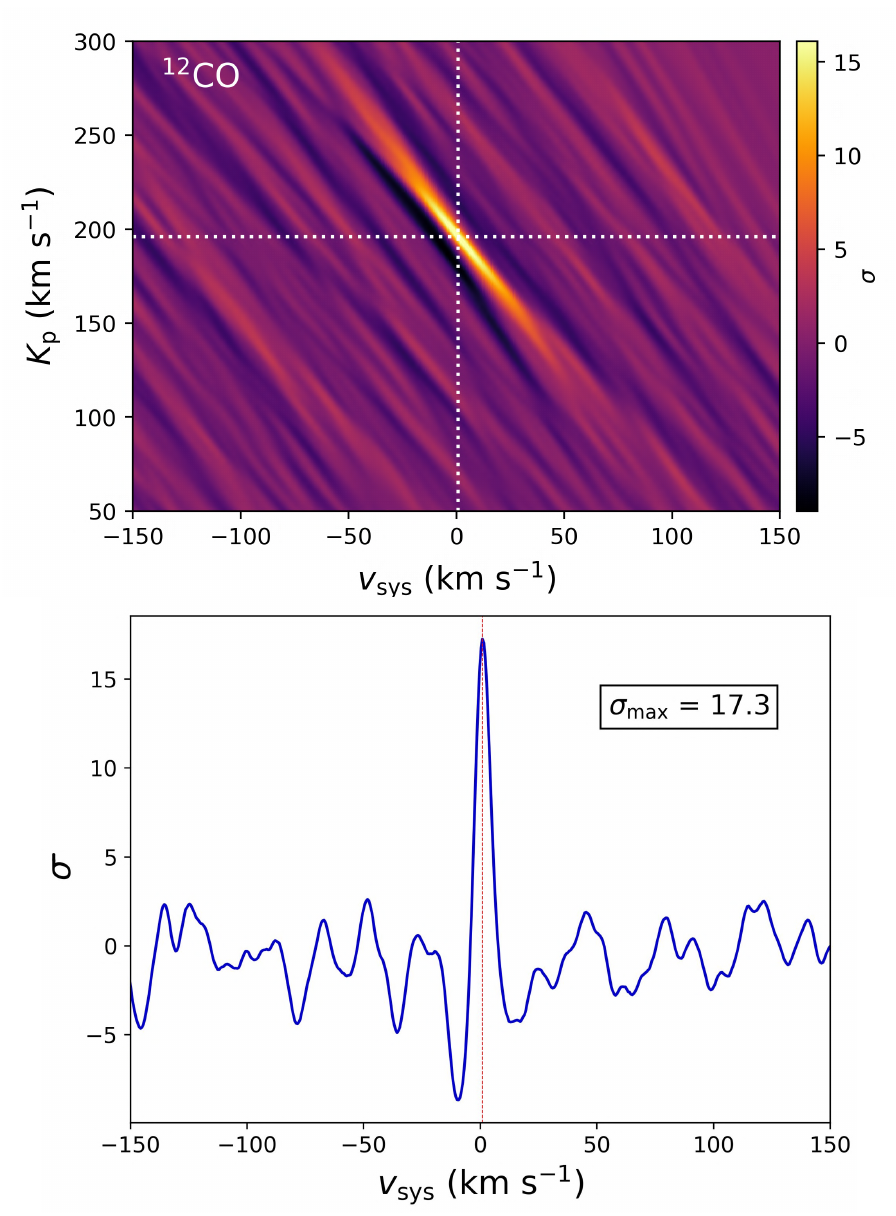}
\end{minipage}%
\hfill 
\begin{minipage}[t]{0.333\textwidth}
  \includegraphics[width=\linewidth]{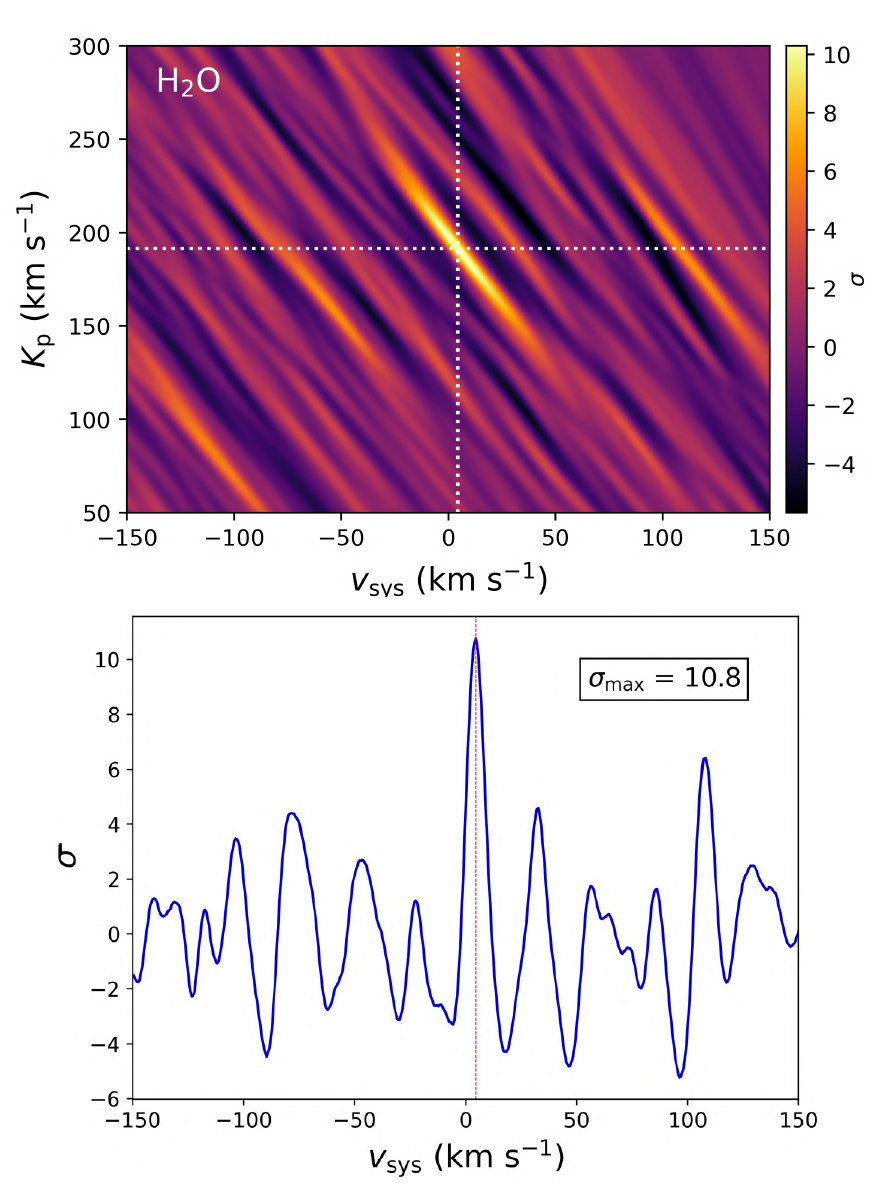}
\end{minipage}%
\hfill
\begin{minipage}[t]{0.333\textwidth}
  \includegraphics[width=\linewidth]{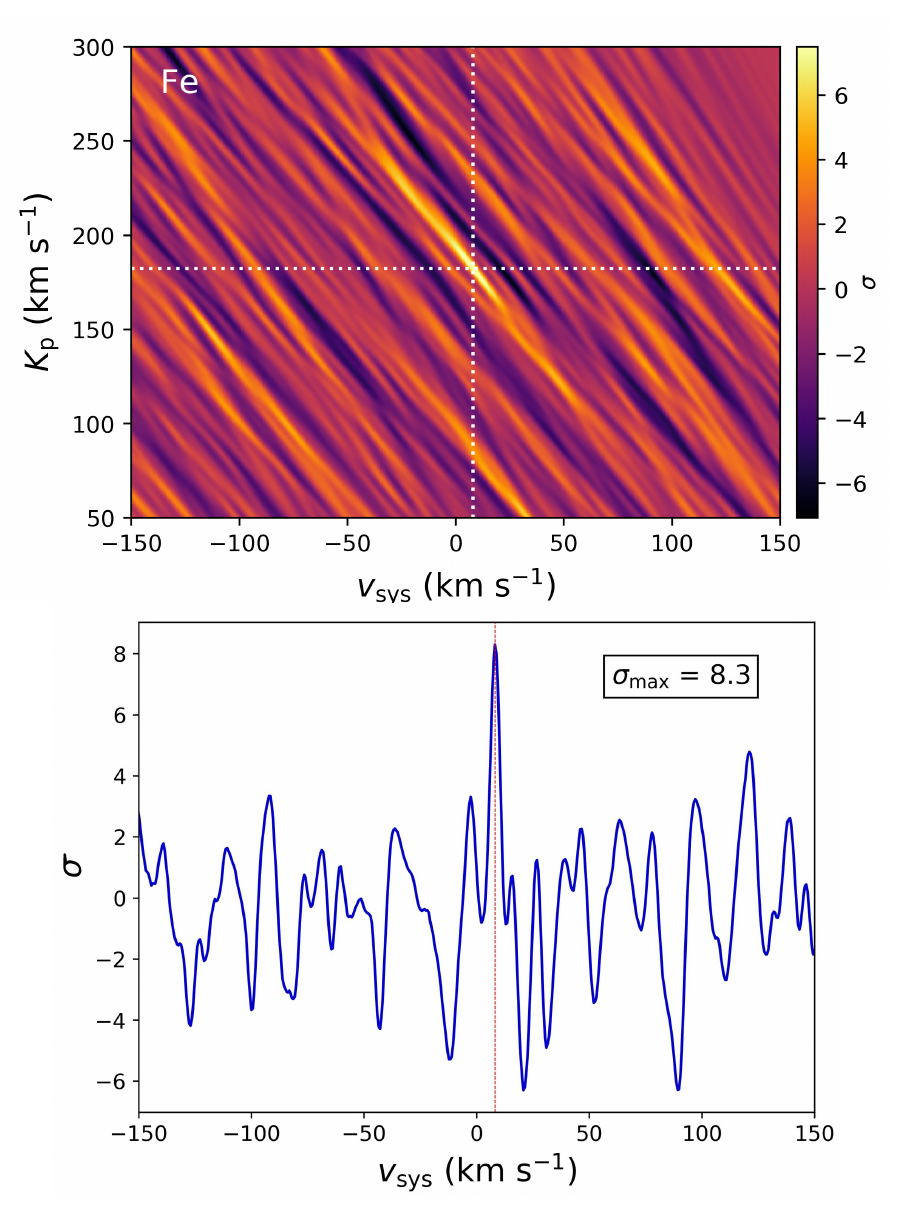}
\end{minipage}%
\caption{Results from cross-correlation for $\rm CO$, $\rm H_2O$ and $\rm Fe$. Top: velocity-summed cross-correlation (\textrm{$K_{\rm p}$-$v_{\rm sys}$}) maps. The white dotted line marks the peak of the detection and the colour bar shows the detection significance weighted on the standard deviation of the cross-correlation map outside the peak, showing a clear detection of $ \mathrm{{}^{12}CO} $, $ \mathrm{H_2O} $, and $ \mathrm{Fe} $ in the atmosphere of MASCARA-1b (see text). \textit{Bottom:} a slice of the \textrm{$K_{\rm p}$-$v_{\rm sys}$} map at peak $K_{\rm p}$.}
\label{fig:kp-vsys_maps}
\end{figure*}
Our forward model atmosphere (for free retrievals), therefore, has $N_\mathrm{species}$ + 4 parameters using the parametric model from \citet{guillot2010radiative} and $N_\mathrm{species}$ + 7 parameters using the model from \citet{madhusudhan2009temperature}, where $N_\mathrm{species}$ refers to the number of atmospheric species under consideration.\\

For our retrieval analysis (detailed in Sect.~\ref{sec:4.2}), we focus on the species that are detected in the day-side atmosphere of MASCARA-1b. Therefore, $N_\mathrm{species}=3$ (namely, $\rm CO$, $\rm H_2O$, and $\rm Fe$). However, we note that this parameter may vary in different MCMC runs as we also include other $\rm C$- and $\rm O$-bearing species, for the sake of completeness, in which case, the value of $N_{\rm species}$ is mentioned within parentheses. The parameter vector is given by \{$\kappa_{\rm IR}$, $\gamma$, $T_{\rm irr}$, $P_{\rm cloud}$, $\chi_{\rm species}$\,$\times$\,$N_{\rm species}$\} and \{$T_0$, $\alpha_1$, $\alpha_2$, $P_1$, $P_2$, $P_3$, $P_{\rm cloud}$, $\chi_{\rm species}$\,$\times$\,$N_{\rm species}$\} for the two \textrm{$T$-$P$} profiles. We used a reference pressure and gravity of $0.01$ bar and 40.76 $ \rm m $ $\rm s^{-2}$, respectively, to correspond to MASCARA-1b’s atmosphere, a mean molecular weight of 2.33 and a stellar radius of $R_*$\,=\,2.1$R_\odot$ \citep{talens2017mascara}. Furthermore, we are fixing $T_{\rm int}$ to be 100 K and $ \alpha $ to be 1 in the model fits.\\

We use uniform prior distributions and divide each spectrum of the forward model by its median value to emulate the blaze correction and filter it as described in Section~\ref{sec:3.3}. The log-posterior is then calculated by adding the log-prior and log-likelihood (Eqn.~\ref{eq:10}) for a given set of model parameters, which is incorporated into an MCMC framework to sample the posterior and obtain an estimate of the posterior distributions of the model parameters. We use a custom Differential-Evolution Markov Chain (DEMC) \citep[e.g.][]{braak2006markov, eastman2013exofast}, running an MCMC chain with 128 walkers, with a burn-in length of 200 and a chain length of 300, resulting in 38,400 samples of the posterior. We test for convergence using the Gelman $\&$ Rubin statistic \citep{gelman1992inference} after splitting the chains into four separate groups. The best-fitting parameters using the filtered model were then used to generate a combined model emission spectra of $\mathrm{CO}$, $ \mathrm{H_2O} $ and $ \mathrm{Fe} $ from which we compute a likelihood map and a conditional likelihood distribution of $ \alpha $ (see Fig.~\ref{fig:bestfit_likelihood_maps}).

\section{Results}\label{sect4}
\subsection{Detection of species}\label{sec:4.1}
Following the procedure outlined in Section~\ref{sec:3.4}, cross-correlation analysis was performed with a filtered model emission spectrum, containing $ \mathrm{{}^{12}CO, \hspace{0.5mm} H_2O} $ and $ \mathrm{Fe} $, using the best-fitting model parameters from our atmospheric retrieval (Sections~\ref{sec:3.6} and~\ref{sec:4.2}). We calculated an orbital velocity-systemic velocity (\textrm{$K_{\rm p}$-$v_{\rm sys}$}) map by shifting the cross-correlation functions (CCFs) to the planetary rest-frame over a range of $K_\mathrm{p}$ from $-300$ to $+300$ $\rm km$ $\rm s^{-1}$, and $v_{\mathrm{sys}}$ from $-300$ to $+300$ $\rm km$ $\rm s^{-1}$, in steps of 0.3 and 0.2 $\rm km$ $\rm s^{-1}$, respectively and summed over time. We note that here, $v_\mathrm{sys}$ is expected to be zero, as the spectra have already been shifted to the stellar rest frame by correcting for both the systemic velocity ($11.20$ $ \rm km $ $\rm s^{-1}$; Table~\ref{tab:example_table1}) and the barycentric velocity. We compute the detection significance by dividing the \textrm{$K_{\rm p}$-$v_{\rm sys}$} map through by its standard deviation taken from a $K_{\rm p}$ of $250$ to $280$ $\rm km$ $\rm s^{-1}$ and a $v_{\rm sys}$ from $-100$ to $-50$ and $50$ to $-100$ $\rm km$ $\rm s^{-1}$, avoiding the peak of the cross-correlation map and zero $K_{\rm p}$.
\begin{figure}
\begin{subfigure}{\columnwidth}
   \includegraphics[width=\columnwidth]{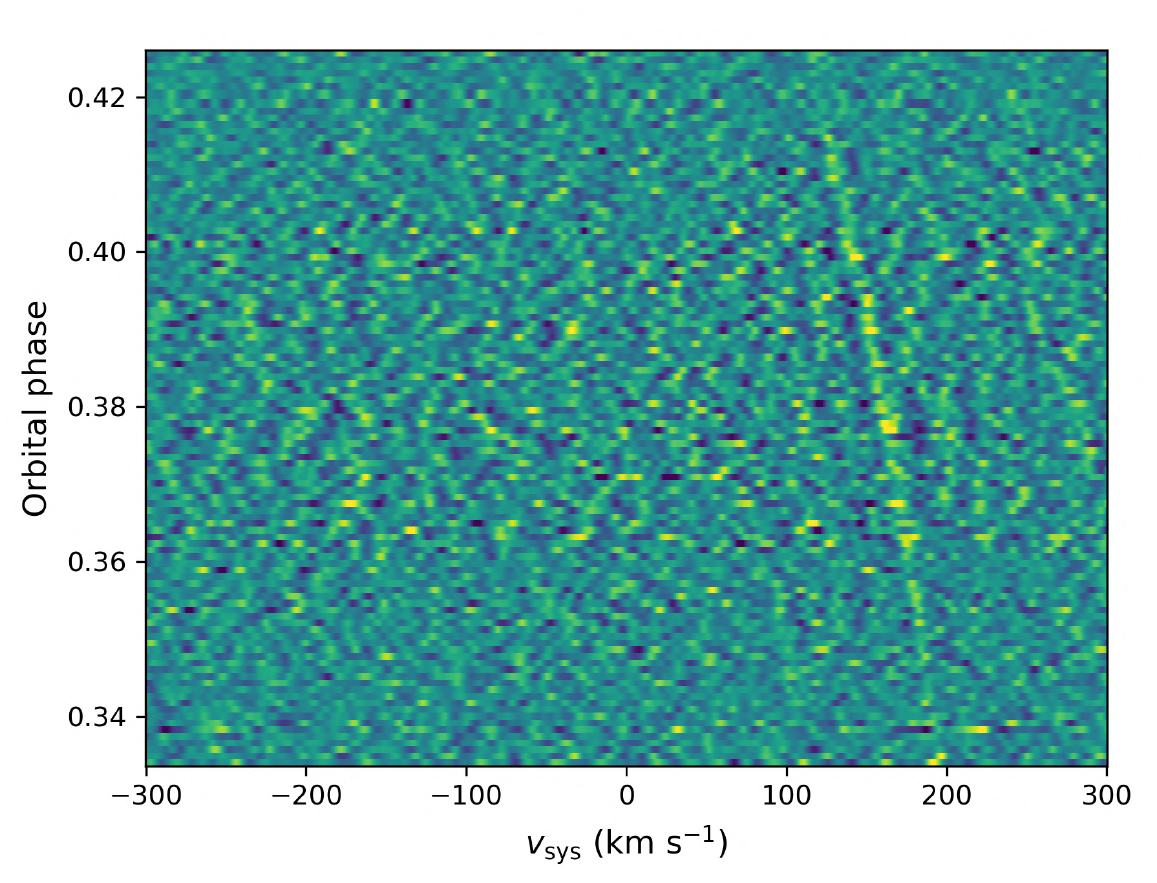}
\end{subfigure}

\begin{subfigure}{\columnwidth}
   \includegraphics[width=\columnwidth]{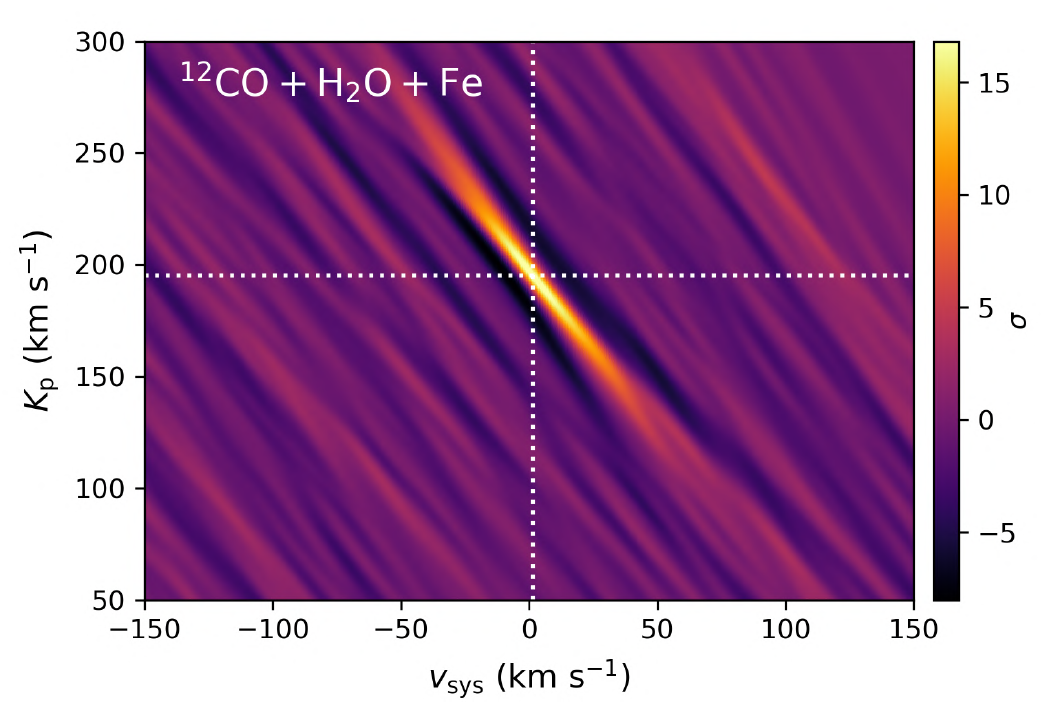}
\end{subfigure}
\caption{Combined detection maps of MASCARA-1b ($ \mathrm{{}^{12}CO} $, $ \mathrm{H_2O} $, and $ \mathrm{Fe} $). The top panel shows the cross-correlation map with the planetary trail visible (slanted feature), and the bottom panel shows the \textrm{$K_{\rm p}$-$v_{\rm sys}$} map. The dotted lines mark the peak of the detection, and the colour bar shows the detection significance (17.8$\sigma$) weighted on the standard deviation of the cross-correlation map outside the peak.}\vspace{14pt}\label{fig:combinedmaps}
\end{figure}
\begin{table}
	\centering
	\caption{Parameters for MASCARA-1 system.\\
    (\textbfss{Notes.} $ ^{(\mathrm{a})} $ \citet{talens2017mascara}, $ ^{(\mathrm{b})} $ \citet{hooton2022spi}, $ ^{(\mathrm{c})} $ \citet{talens2017mascara} found an offset in the systemic velocities derived from two different data sets.)}
	\label{tab:example_table1}
    \renewcommand{\arraystretch}{1.3} 
	\begin{tabular}{ccc} 
		\hline
		Parameter & Symbol & Value\\
		\hline
		Effective temperature [K] & $ \mathrm{T_{eff}} $ & $ 7554\pm 150 $ $ ^\mathrm{a} $\\
        Metallicity & $\rm [Fe/H]$ & $0$ $ ^\mathrm{a} $\\
		Stellar mass & $ \mathrm{M_*} $ & $ 1.72\pm 0.07 \hspace{0.1cm} \textup{M}_\odot\ $ $ ^\mathrm{a} $\\
		Stellar radius & $ \mathrm{R_*} $ & $ 2.1\pm 0.2 \hspace{0.1cm} \textup{R}_\odot\ $ $ ^\mathrm{a} $\\
        Planet mass & $ \mathrm{M_p} $ & $ 3.7 \pm 0.9 \hspace{0.1cm} \textup{M}_\mathrm{jup} $ $ ^\mathrm{a} $\\
        Planet radius & $ \mathrm{R_p} $ & $ 1.5 \pm 0.3 \hspace{0.1cm} \textup{R}_\mathrm{jup} $ $ ^\mathrm{a} $\\
        Equilibrium temperature [K] & $ \mathrm{T_{eq}} $ & $ 2570^{+50}_{-30} \hspace{0.2cm} $ $ ^\mathrm{a} $\\
        Epoch [BJD] & $ \mathrm{T_p} $ & $ 2457097.278 \pm 0.002 $ $ ^\mathrm{a} $\\
        Period [days] & $ \mathrm{P} $ & $ 2.148780 \pm 0.000008 $ $ ^\mathrm{a} $\\
        Semi-major axis [AU] & $ a $ & $ 0.043 \pm 0.005 $ $ ^\mathrm{a} $\\
        Eccentricity & $ e $ & $ 0 $ (fixed) $ ^\mathrm{a} $\\
        Inclination & $ i $ & $ 87^{\circ^{+2}}_{-3} $ $ ^\mathrm{a} $\\
        RV semi-amplitude [$ \rm km $ $\rm s^{-1}$] & $ \mathrm{K_p} $ & $ 217 \pm 25 $ $ ^\mathrm{a} $\\
        & & $ 204.2 \pm 0.2 $ $ ^\mathrm{b} $\\
        Systemic velocity [$ \rm km $ $\rm s^{-1}$] & $ \gamma, \mathrm{\mathrm{v_{sys}}} $ & $ 11.20 \pm 0.08 $; $ 8.52 \pm 0.02 $ $ ^\mathrm{c} $\\
		\hline
	\end{tabular}
\end{table}
From the cross-correlation analysis, we find strong emission signatures of $ \mathrm{{}^{12}CO} $, $ \mathrm{H_2O} $, and $ \mathrm{Fe} $ in the day-side atmosphere of MASCARA-1b. We detect $\rm CO$, $\rm H_2O$ and $\rm Fe$ at roughly the expected $K_{\rm p}$ \citep[${\approx}217\pm25$ $\rm km$ $\rm s^{-1}$ from][]{talens2017mascara} and $v_{\rm sys}$ (0 $\rm km$ $\rm s^{-1}$) with a detection significance of 17.3$\sigma$, 10.8$\sigma$ and 8.3$\sigma$, respectively. The \textrm{$K_{\rm p}$-$v_{\rm sys}$} maps for individual species are shown in Fig.~\ref{fig:kp-vsys_maps}. Using our combined model with all three species, we obtain a significance of $17.8$$\sigma$. The combined cross-correlation and \textrm{$K_{\rm p}$-$v_{\rm sys}$} maps are shown in Fig.~\ref{fig:combinedmaps}.\\

We note that atomic species (e.g. $ \mathrm{Fe} \hspace{0.3mm} \textsc{i} $, $ \mathrm{Fe} \hspace{0.3mm} \textsc{ii} $, $ \mathrm{TiO} $, $ \mathrm{Ca} \hspace{0.3mm} \textsc{i} $, etc) were previously searched for in the atmosphere of MASCARA-1b using high-resolution transmission spectroscopy with HARPS \citep{stangret2022high} and ESPRESSO \citep{casasayas2022transmission}. Both analyses reported non-detection of absorption features due to the presence of a strong Rossiter-McLaughlin (RM) effect, causing an overlap of any potential planetary signal with the Doppler shadow (prominent for $ \mathrm{Fe} \hspace{0.3mm} \textsc{i} $, $ \mathrm{Fe} \hspace{0.3mm} \textsc{ii} $, $ \mathrm{Ca} \hspace{0.3mm} \textsc{i} $). However, recent high-resolution detections of $\rm CO$ and $\rm H_2O$ with CRIRES+ \citep{holmberg2022first} as well as $\rm Ti \hspace{0.3mm} \textsc{i}$, $\rm Cr \hspace{0.3mm} \textsc{i}$ and $\rm Fe \hspace{0.3mm} \textsc{i}$ with PEPSI \citep{2023arXiv230403328S} show that the atmosphere of MASCARA-1b can be detected through {\it emission} spectroscopy as it does not suffer from any overlapping RM effect, thus allowing us to detect $\rm Fe$ ($\approx$8$\sigma$) in the K-band. We also confirm the previously reported detections of $ \mathrm{CO} $ and $ \mathrm{H_2O} $ in the day-side atmosphere of MASCARA-1b by \citet{holmberg2022first} which are consistent with our reported values. We note that our data points are weighted by variance before summing over wavelength to generate our optimal $ \mathrm{CCF} $ as a function of time/phase and $ v_\mathrm{sys} $, therefore resulting in a higher significance for our detection. If we do not weight by the individual uncertainties (equivalent to assuming identical uncertainties for all times and wavelengths), the detection significance drops to ${\approx}11\sigma$. This highlights the importance of fully taking into account the heteroskedastic nature of the noise.

\subsection{Retrievals}\label{sec:4.2}
\begin{figure*}
  \includegraphics[width=1.93\columnwidth]{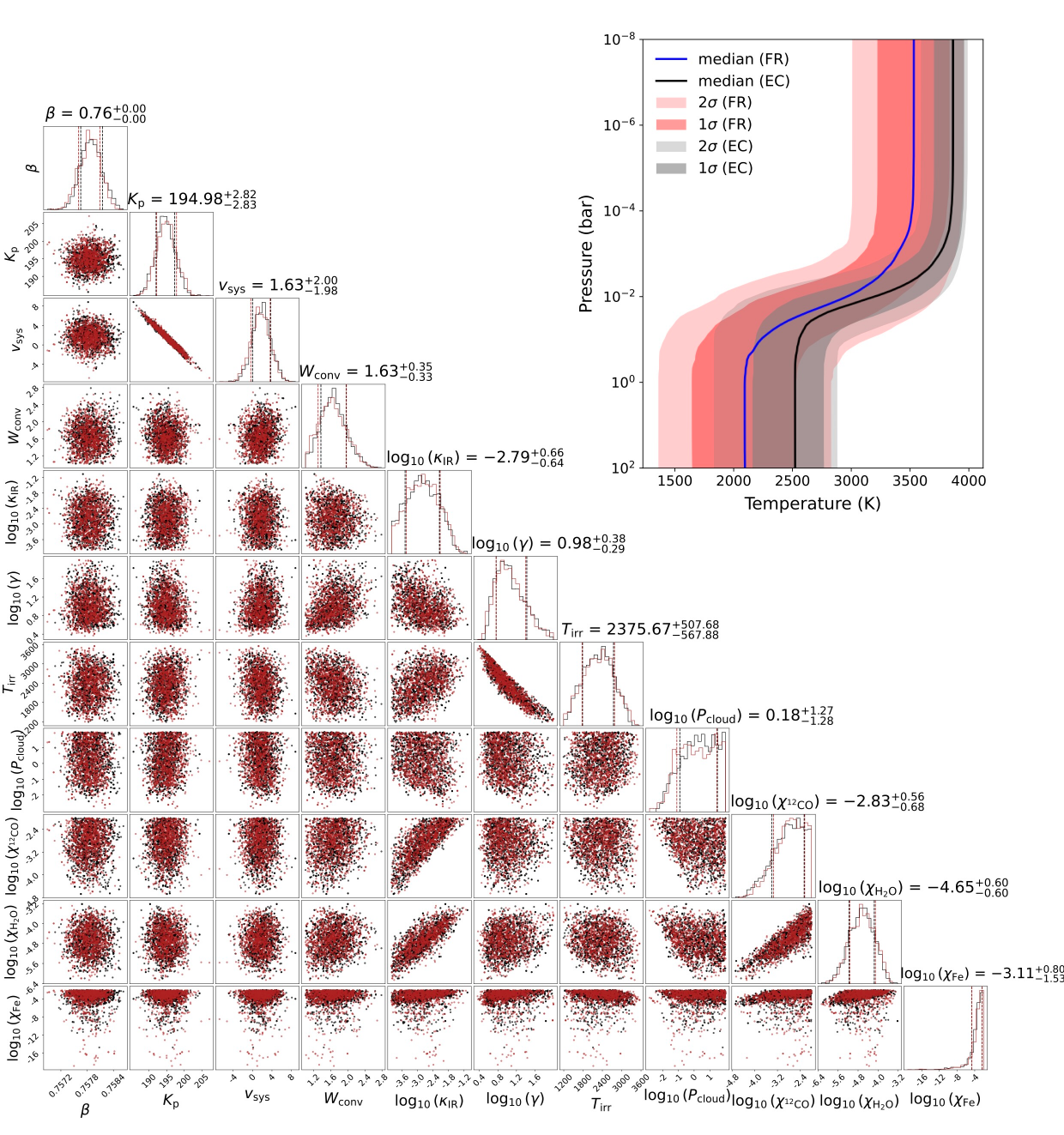}
  \caption{A summary of our `free-retrieval' results for MASCARA-1b with the 1D and 2D marginalised posterior distributions of each model parameter displayed within a corner plot. The red and black posterior distributions represent independent sub-chains of the same MCMC chain, both converging to similar distributions. \textit{Upper right:} the parametric \textrm{$T$-$P$} profile \citep{guillot2010radiative} computed from the best-fitting model parameters. The blue and black solid lines are the median profiles for the `free-retrieval' (labelled FR) and the chemistry (labelled EC) frameworks. The red and grey shading marks the $ 1\sigma $ and $ 2\sigma $ recovered distribution computed from 10,000 samples from the MCMC for both regimes.}\label{fig:chemmodel_guillot}
\end{figure*}

Computing full log-likelihood maps from the CCFs as a function of $K_\mathrm{p}$ and $v_\mathrm{sys}$, as well as model parameters of the atmospheric model and likelihood quickly becomes prohibitive as the number of parameters increases. Rather than effectively compute the log-likelihood for a grid of parameters, it is much more efficient to use Markov Chain Monte Carlo (MCMC) techniques. Following the procedure discussed in Section~\ref{sec:3.6}, we begin the retrieval process with the basic `free-retrieval' paradigm, which assumes constant-with-altitude mixing ratios and uses the parametric model from \citet{guillot2010radiative}. The results are shown in Fig.~\ref{fig:chemmodel_guillot} and the marginalised distributions for each parameter are summarised in Table.~\ref{retrievalpars_guillot}.\\

As mentioned in Section~\ref{sec:3.2}, the ratio of visible-to-infrared opacity $\gamma$ governs the atmospheric temperature gradient. Values of $\gamma$\,=\,1 produce isothermal atmospheres; $\gamma$\,<\,1 produce decreasing temperatures with decreasing pressure; and $\gamma$\,>\,1 result in temperature inversions (increasing temperatures with decreasing pressure).
The retrieved $ \log_{10}(\gamma) $ for MASCARA-1b is $ 0.98^{+0.40}_{-0.28} $ which is a confirmation of a \textrm{$T$-$P$} profile with thermal inversion (see Fig.~\ref{fig:chemmodel_guillot}). The retrieved $ \mathrm{{}^{12}CO} $ abundance from our free-retrieval setup for MASCARA-1b is $ \log_{{10}}(\chi_{\mathrm{CO}}) = -2.85^{+0.57}_{-0.69} $ which is suggestive of a carbon-rich atmosphere. Here we infer the free-retrieval based $\mathrm{C/O}$ and metallicity of MASCARA-1b by counting the total elemental number density arising from each species. The solar elemental abundances are taken from \citet{asplund2009chemical}. We compute the planetary $ \mathrm{C/O} $ ratio as follows:
\begin{equation*}
    \mathrm{C/O} = \frac{n_\mathrm{C}}{n_\mathrm{O}} = \frac{n_\mathrm{CO}}{n_\mathrm{CO} + n_\mathrm{H_2O}}
\end{equation*}
Here, we assume that $\mathrm{{}^{12}CO}$ and $\mathrm{H_2O}$ are the dominant carbon- and oxygen-bearing molecules in the atmosphere of MASCARA-1b, and we compute the planetary $\rm M/H$ by normalising the elemental abundances relative to hydrogen ($\rm n_i/n_H$), relative to that in the Sun ($\rm [X/H]$\,=\,$\log_{10}((\rm n_X/n_H)/(\rm n_X/n_H)_\odot)$).\\

We find the elemental abundances in the atmosphere of MASCARA-1b to be\footnote{Square brackets (`[]') refers to the $\log_{10}$ abundances relative to solar.} $\rm [C/H]$\,=\,$0.46^{+0.57}_{-0.69}$ (0.6-10\,$\times$\,solar), $\rm [O/H]$\,=\,$0.21^{+0.56}_{-0.69}$ (0.3-5.8\,$\times$\,solar), $\rm [Fe/H]$\,=\,$0.97^{+0.81}_{-1.52}$ (0.3-60\,$\times$\,solar) (MASCARA-1 has been measured to have a solar $\rm [Fe/H]$; see Table~\ref{tab:example_table1}), and a $ \mathrm{C/O} = 0.98^{+0.01}_{-0.02} $ which is super-solar. The elevated $ \mathrm{CO} $ abundance relative to $\mathrm{H_2O}$ drives the $\rm C/O$ ratio towards 1 as well as results in an unrealistically small uncertainty using this method.\\
\\
Overall, our free-retrieval derived abundances are indicative of a planetary atmosphere that is super-solar in $ \mathrm{C/O} $. We also re-ran our free-chemistry retrieval to include potential $\rm C$- and $\rm O$-bearing species that were not detected (e.g. $ \mathrm{OH} $, $ \mathrm{CO_2} $, $ \mathrm{HCN} $, and $ \mathrm{CH_4} $; here $N_{\rm species} = 7$), and find that the inclusion of these species does not change the retrieved $\rm C/O$, which remains super-solar ($0.98^{+0.01}_{-0.02}$; see Fig.~\ref{fig:freeret_allspecies}). There are, however, several shortcomings within the free-retrieval setup, which assumes constant vertical abundances for the species, that might lead to biases -- particularly for UHJs.
\begin{table*}
  \centering
  \caption{Parameters recovered for the combined fits of MASCARA-1b for two different chemical regimes using the parametric $T$-$P$ profile from \citet{guillot2010radiative}.}
  \label{retrievalpars_guillot}
  \renewcommand{\arraystretch}{1.37} 
  \begin{tabular}{ccccc}
    \hline
    Parameter [units] & Prior & Free-retrieval & Equilibrium chemistry\\
    \hline
     $ \alpha $ & - & - & - \\
     $ \beta $ & $ \mathcal{U}(0.1, 2) $ & $ 0.76 \pm 0.0003 $ & $ 0.76 \pm 0.0003$ \\
     $ K_\mathrm{p} $ [$ \rm km $ $\rm s^{-1}$] & $ \mathcal{U}(185, 215) $ & $ 194.7^{+2.8}_{-2.7} $ & $ 194.6 \pm 2.90 $ \\
     $ v_\mathrm{sys} $ [$ \rm km $ $\rm s^{-1}$] & $ \mathcal{U}(-15, 15) $ & $ 1.77 \pm 1.90 $ & $ 1.86 \pm 2.00$ \\
     $ W_\mathrm{conv} $ & $ \mathcal{U}(1, 50) $ & $ 1.65^{+0.33}_{-0.31} $ & $ 1.54^{+0.35}_{-0.29} $ \\
     $ \log_{{10}}(\kappa_\mathrm{IR}) $ [$ \rm m^{2} $ $\rm kg^{-1}$] & $ \mathcal{U}(-4, 0) $ & $ -2.80^{+0.67}_{-0.65} $ & $ -2.40^{+0.33}_{-0.47} $ \\
     $ \log_{{10}}(\gamma) $ & $ \mathcal{U}(-2, 2) $ & $ 0.98^{+0.40}_{-0.28} $ & $ 0.81^{+0.30}_{-0.17} $ \\
     $ T_\mathrm{irr} $ [$ \mathrm{K} $] & $ \mathcal{U}(1000, 4000) $ & $ 2367^{+502}_{-549} $ & $ 2843^{+273}_{-456} $ \\
     $ T_\mathrm{int} $ [$ \mathrm{K} $] & - & - & -\\
     $ \log_{{10}}(P_\mathrm{cl}) $ [bar] & $ \mathcal{U}(-4, 2) $ & $ 0.26^{+1.20}_{-1.28} $ & $ 0.18^{+1.21}_{-1.29} $ \\
     $ \log_{{10}}(\chi_{\rm CO}) $ & $ \mathcal{U}(-20, -2) $ & $ -2.85^{+0.57}_{-0.69} $ & - \\
     $ \log_{{10}}(\chi_{\rm H_2O}) $ & $ \mathcal{U}(-20, -2) $ & $ -4.66^{+0.58}_{-0.60} $ & - \\
     $ \log_{{10}}(\chi_{\rm Fe}) $ & $ \mathcal{U}(-20, -2) $ & $ -3.13^{+0.81}_{-1.52} $ & - \\
     $ [\mathrm{M/H}] $ & $ \mathcal{U}(-1, 1) $ & $ 0.52^{+0.44}_{-0.72} $ & $ 0.62^{+0.28}_{-0.55} $ \\
     $ \log_{{10}}(\mathrm{C/O}) $ & $ \mathcal{U}(-1, 1) $ & $ -0.011^{+0.00}_{-0.01} $ & $ -0.17^{+0.08}_{-0.17} $ \\
    \hline
  \end{tabular}
\end{table*}
\begin{figure*}
  \includegraphics[width=1.306\columnwidth]{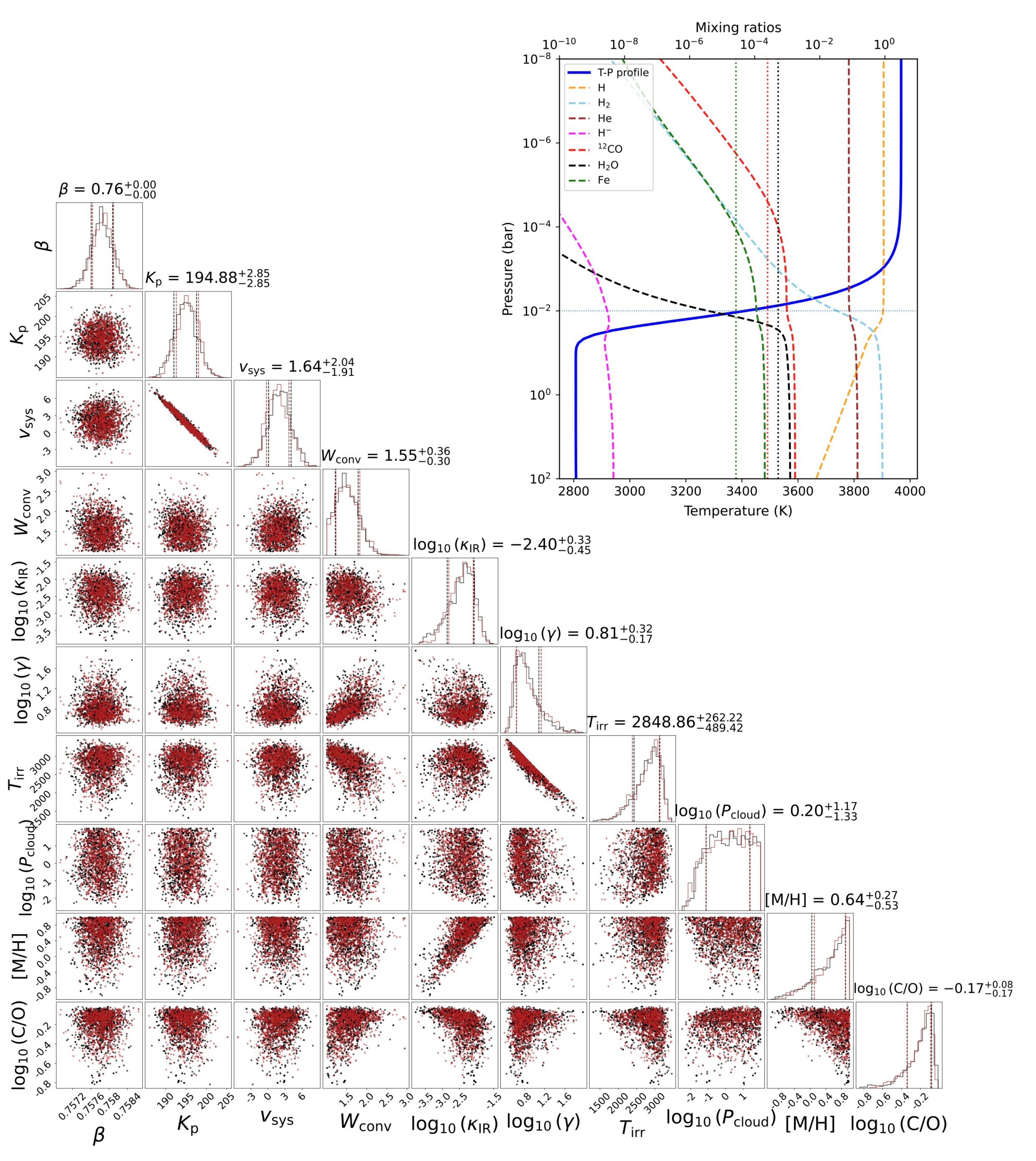}
  \caption{A summary of our equilibrium chemistry retrieval results for MASCARA-1b with the 1D and 2D marginalised posterior distributions of each model parameter displayed within a corner plot. \textit{Upper right:} the atmospheric structure from the best-fitting chemistry model. The volume mixing ratio profiles for continuum species and the detected species are shown as dashed lines (calculated using \textsc{FastChem}). The parametric $T$-$P$ profile \citep{guillot2010radiative} is shown as a solid blue line and constant mixing ratios for the detected species as dotted lines.}\label{fig:chem_ret_tpenvelope}
\end{figure*}
\begin{figure*}
  \includegraphics[width=1.89\columnwidth]{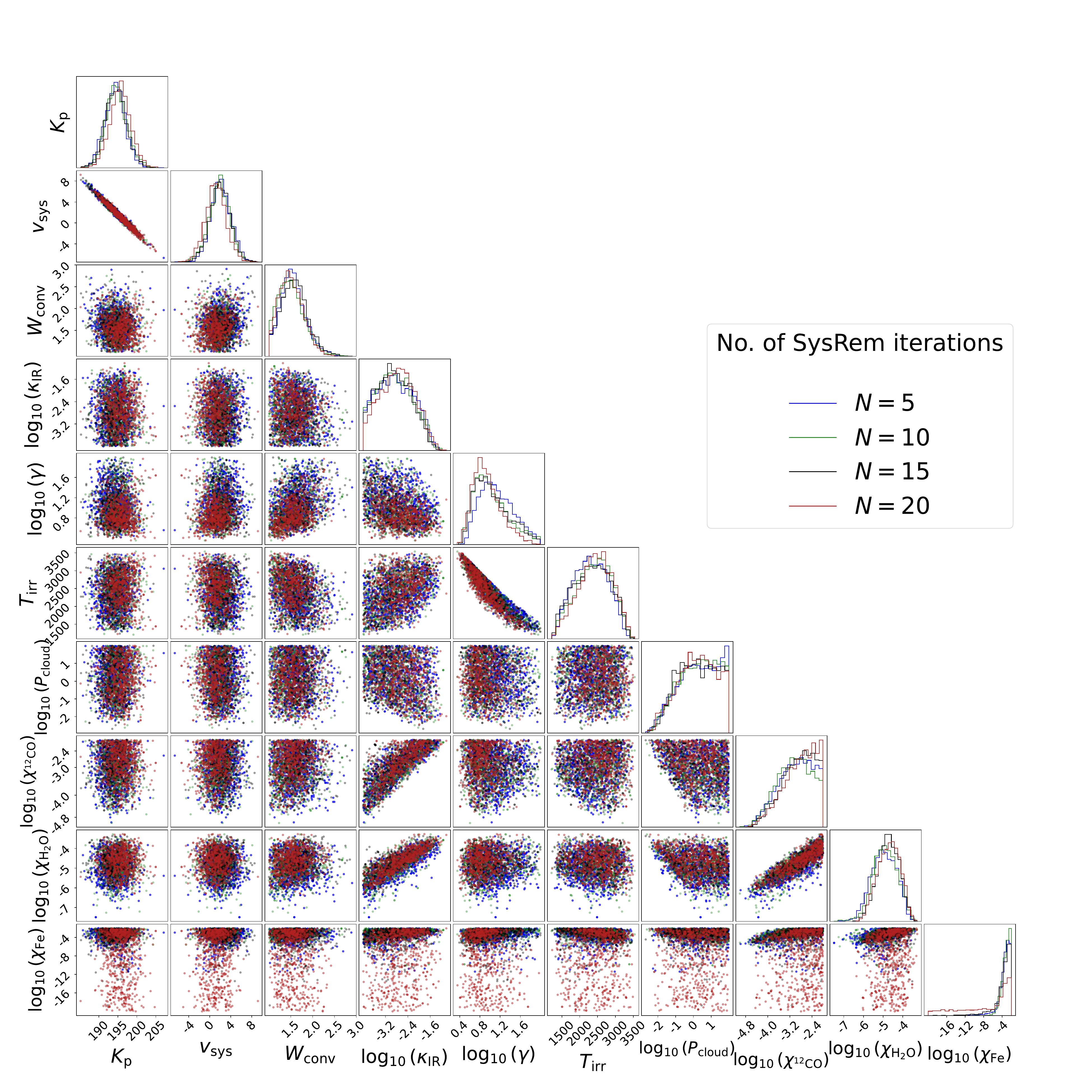}
  \caption{A summary of our free-retrieval results for MASCARA-1b with the 1D and 2D marginalised posterior distributions of the parameters from the MCMC fit. The different colours show the samples obtained using four different values for the \textsc{SysRem} iterations.}\label{fig:sysrem_iters}
\end{figure*}
We note that for such a framework, inclusion or exclusion of H$_2$-H$_2$ and H$_2$-He collision-induced absorption (CIA) does not affect the $\rm C/O$ ratio, which remains $ 0.98^{+0.01}_{-0.02} $. Whereas, for our retrievals assuming chemical equilibrium, we find that the $\rm{H_2}$ and $\rm{H}$ abundances change quite drastically with altitude in this temperature regime (see upper right panel of Fig.~\ref{fig:chem_ret_tpenvelope}). Therefore, instead of retrieving for the individual gas volume mixing ratios, we used \textsc{FastChem} \citep{stock2018fastchem} to estimate the abundances of chemical species and fit directly for the metallicity and $ \mathrm{C/O} $ ratio. The results of this analysis are shown in Fig.~\ref{fig:chem_ret_tpenvelope} and the marginalised distributions for each parameter are summarised in Table.~\ref{retrievalpars_guillot}. This setup results in a $\rm C/O$ ratio of $ 0.68^{+0.12}_{-0.22} $ and a metallicity of $ \mathrm{[M/H]} = 0.62^{+0.28}_{-0.55} $, both consistent with solar values within $\approx$1.1$\sigma$. The retrieval frameworks were run using the four-parameter $T$-$P$ profile from \citet{guillot2010radiative}. Likewise, we re-compute our forward model using the six-parameter profile from \citet{madhusudhan2009temperature} and perform retrievals assuming constant vertical abundances as well as chemical equilibrium.

A free-retrieval results in $ \log_{10}(\chi_\mathrm{{}^{12}CO}) = -2.66^{+0.44}_{-0.67} $ and a super-solar $\rm C/O$ of $ 0.98^{+0.01}_{-0.02} $, whereas a chemical retrieval results in a $\rm C/O$ of $ 0.48^{+0.29}_{-0.24} $ and $\rm [M/H] = 0.67^{+0.24}_{-0.40}$. The results of this analysis are shown in Figures~\ref{fig:chemmodel_mstp}, ~\ref{fig:chem_ret_mstp_chemmodel}, and~\ref{fig:chem_ret_mstp_tpenvelope}, and a summary of retrieved parameter values are outlined in Table.~\ref{retrievalpars_mstp}. We also drew 10,000 random samples from each of the MCMC and computed the $T$-$P$ profile for both retrieval frameworks using two different parametrizations of the $T$-$P$ profile \citep{guillot2010radiative, madhusudhan2009temperature}, and find them to be consistent (see upper right panel of Fig.~\ref{fig:chemmodel_guillot} and Fig.~\ref{fig:chemmodel_mstp}). Additionally, we also test the effect of the number of \textsc{SysRem} iterations on our retrievals. As noted in Section~\ref{sec:2.2}, we use an arbitrary number of SysRem iterations for our retrievals ($N = 15$). Therefore, we re-ran our framework with $ N = 5, 10 $ and $20$ iterations and find that using $N = 5 $, $N = 10$, and $N = 15$ give us consistent results; however, $N = 20$ filtered out the $\rm{Fe}$ signal. Since \textsc{SysRem} removes parts of each signal (on increasing the number of iterations), and because the $\rm Fe$ signal is already weaker compared to $\rm CO$ and $\rm H_2O$, it is not constrained. The results of this analysis are shown in Fig.~\ref{fig:sysrem_iters}, which highlights the fact that, for MASCARA-1b, increasing the number of \textsc{SysRem} iterations can eventually filter out the exoplanet signal, resulting in a loss of information. In summary, we perform retrievals using four different types of models: two parametric $T$-$P$ profiles and two different chemical regimes.

\section{Discussion}\label{sect5}
We present the first retrieval results for the ultra-hot Jupiter, MASCARA-1b, with the upgraded CRIRES+, using observations from the science verification run. Our results demonstrate clear detections of $\mathrm{CO}$, $\mathrm{H_2O}$, and $\mathrm{Fe}$ (first reported detection in the K-band) in the day-side atmosphere of MASCARA-1b. Using standard cross-correlation analysis, we find a detection significance of $ 17.3\sigma $, $ 10.8\sigma $, and $ 8.3\sigma $, for $ \mathrm{CO} $, $ \mathrm{H_2O} $, and $ \mathrm{Fe} $, respectively. Through the emission features of $ \mathrm{CO}$, $ \mathrm{H_2O} $, and $ \mathrm{Fe} $, we also confirm the presence of a thermal inversion layer in the atmosphere of MASCARA-1b. The cross-correlation value can also be mapped to a likelihood value, as outlined in Section~\ref{sec:3.5}, and a direct likelihood evaluation then enables a full retrieval framework. Thus, we may constrain the absolute abundances of each species, as well as the velocity shifts, $T$-$P$ structure, $\rm C/O$ ratio, etc.\\

\begin{figure}
  \includegraphics[width=0.85\columnwidth]{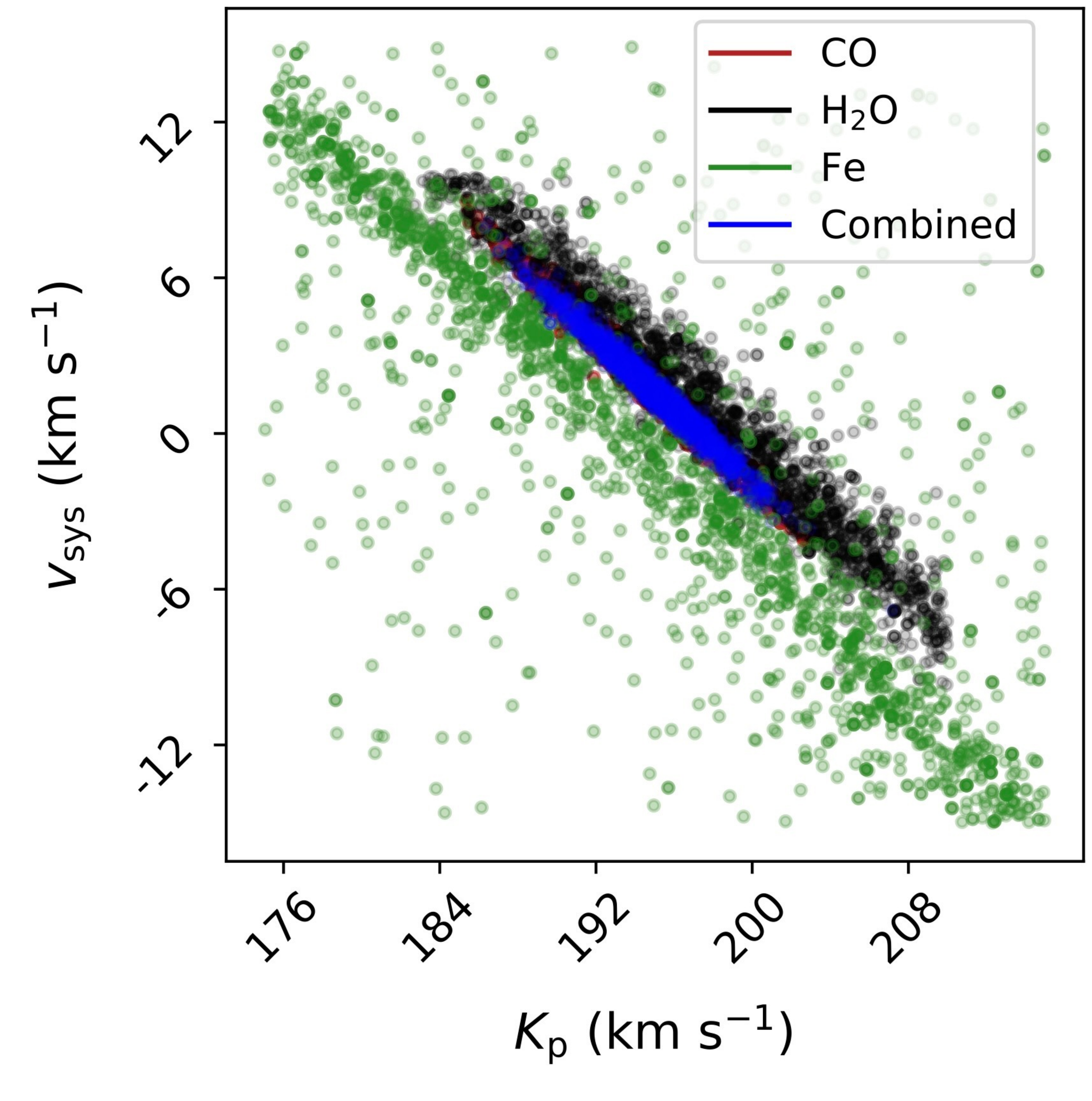}
  \caption{The 2D marginalised posterior distribution for $\rm CO$, $\rm H_2O$, $\rm Fe$ and their combined spectrum show velocity offset for the $\rm Fe$ feature (see text).}\label{fig:offsets}
\end{figure}

Our observations detect a slight offset of the $ \mathrm{Fe} $ feature in both $ K_\mathrm{p} $ and $ v_\mathrm{sys} $ (Section~\ref{sec:4.1}), and the 2D corner plots (see Fig.~\ref{fig:offsets}) show a strong hint that the signals are separated in velocity space. This is apparent in the CCF as well as the MCMC fits, and we exclude that the measured shifts are due to inaccurate line positions as per the analysis of \citet{gandhi2020molecular} who found that the line lists of $\mathrm{CO}$ and $\mathrm{H_2O}$ are appropriate for high-resolution studies up to $R$\,=\,$100{,}000$. We also note that the shifts are measured in both $K_\mathrm{p}$ and $v_\mathrm{sys}$ (similar to the shifts detected for $\rm CO$ and $\rm H_2O$ by \citealt{brogi2023roasting} in WASP-18b). These could be due to different altitudes probed by different species, to their emission arising from different parts of the planet's atmosphere, or a combination of both, which stresses the fact that exoplanet atmospheres are 3D structures. Further analysis is needed to confirm or refute this result.\\

For our retrieval framework, we employ two different approximations for the description of the atmosphere's chemical composition: a free-retrieval of the mixing ratios for the species and an equilibrium chemistry model \citep[\textsc{FastChem;}][]{stock2018fastchem} to self-consistently calculate the abundances. The free-retrieval setup assumes constant-with-altitude volume mixing ratios for the species. Therefore, the retrieved $ \mathrm{C/O} $ ratio is only constrained by our detection of $ \mathrm{CO} $ and $ \mathrm{H_2O} $, assuming that every bit of $\mathrm{C}$ and $\mathrm{O}$ is within $ \mathrm{CO} $ and $\mathrm{H_2O}$. Such a model produces a super-solar $\rm C/O$ ratio of $ 0.98^{+0.01}_{-0.02} $ which is driven by an elevated $ \mathrm{CO} $ abundance (see Table~\ref{retrievalpars_guillot} and Fig.~\ref{fig:chemmodel_guillot}). However, there could be a significant amount of $\mathrm{C}$ and/or $\mathrm{O}$ in $\mathrm{CO_2}$, $\mathrm{OH}$, $\mathrm{CH_4}$, etc. Thus, we updated our forward model to incorporate $\mathrm{CO_2}$, $\mathrm{OH}$, $\mathrm{CH_4}$, $ \mathrm{HCN} $ and re-ran our retrievals.
\begin{figure}
  \includegraphics[width=0.997\columnwidth]{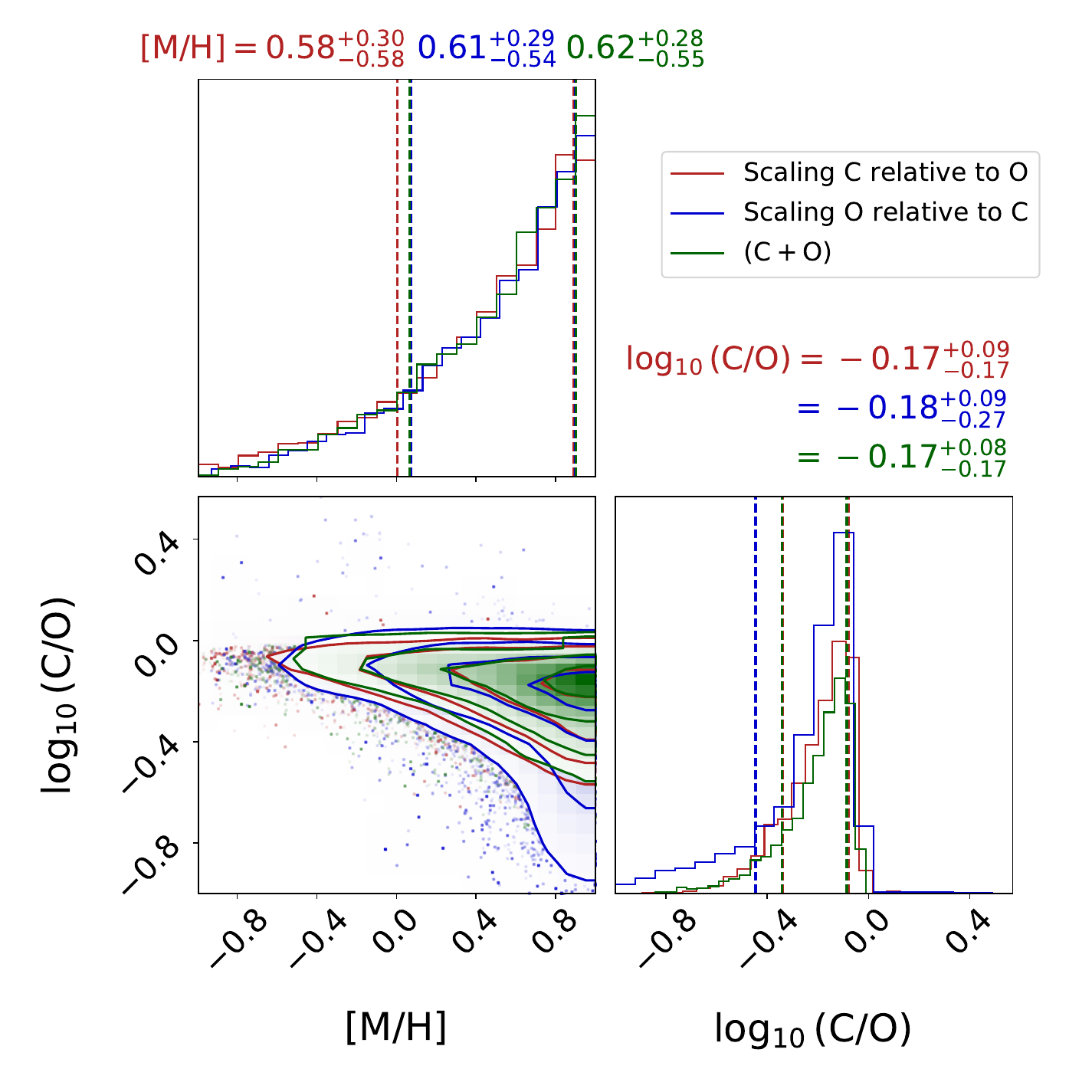}
  \caption{Retrieved values for the carbon-to-oxygen ($\rm C/O$) ratio and metallicity [$\rm M/H$] using the parametric profile from \citet{guillot2010radiative}, assuming chemical equilibrium. The different colours represent how the $\rm C/O$ ratio is adjusted.}\label{fig:Scaling_CtoO}
\end{figure}
We find that the inclusion of potential carbon and oxygen-bearing species to our free-retrieval setup has no impact on the retrieved $\mathrm{C/O}$ ratio, which remains super-solar ($0.98^{+0.01}_{-0.02}$; see Fig.~\ref{fig:freeret_allspecies}).\\

Overall, our free-retrieval-based results strongly favoured a $\rm H_2O$-depleted model over one with a similar abundance of $\rm CO$ and $\rm H_2O$, thereby driving the $\rm C/O$ towards $1$ resulting in precise constraints. This could, however, lead to biases in ultra-hot Jupiters as it assumes constant vertical abundances for the species. Therefore, instead of retrieving for the individual VMRs, we fit directly for the $\rm C/O$ ratio and metallicity derived from the chemical equilibrium model fits. We find that such a model results in a $\rm C/O$ ratio of $0.71^{+0.14}_{-0.21}$ and a metallicity, [$\rm M/H$]\,=\,$0.59^{+0.31}_{-0.53}$, both consistent with solar values within $\approx$1$\sigma$ (MASCARA-1 has been measured to have a solar $[\rm Fe/H]$; see Table~\ref{tab:example_table1}). The results of this analysis are shown in Fig.~\ref{fig:chemret_allspecies}. We note that the exclusion of $\rm{H^-}$ opacity and the collision-induced absorption of H$_2$-H$_2$ and H$_2$-He from our chemical retrievals did not significantly impact the results and the $\rm C/O$ ($ 0.63^{+0.18}_{-0.35} $) remains to be consistent with the solar value. In addition, we test the effect of adjusting the $\rm C/O$ ratio on our retrievals and note that for our analysis assuming chemical equilibrium, the $\rm C/O$ adjusts the relative $\rm C$ \textit{and} $\rm O$ while preserving their sum after setting the abundances with respect to the metallicity. We also tried an alternative by scaling $\rm O$ relative to $\rm C$ and by scaling $\rm C$ relative to $\rm O$ and find that the choice of scaling does not make a difference to our retrievals and the retrieved $\rm C/O$ ratios are consistent (see Fig.~\ref{fig:Scaling_CtoO}). In summary, we use four different models: two parametric $T$-$P$ profiles and two chemical regimes, to perform atmospheric retrievals and constrain the metallicity and $\rm C/O$ ratio of MASCARA-1b's atmosphere. The $ \mathrm{C/O} $ ratio in a planet potentially provides critical information about its primordial origins and subsequent evolution and is also predicted to regulate the atmospheric chemistry in hot/ultra-hot Jupiters \citep{oberg2011effects}. While a high $ \mathrm{C/O} $ ($\sim$1) found with our free retrieval is intriguing, our reported abundance constraints are likely biased due to strong vertically-changing chemical profiles. Our chemical equilibrium retrievals provide more realistic and conservative constraints and are consistent with solar values.\\

We note that MASCARA-1b is only the second UHJ after WASP-18b \citep{brogi2023roasting}, where high-resolution spectroscopy is revealing the limits of free-chemistry, constant-with-altitude abundance modelling. Therefore, similar to their analysis, we highlight the importance of accounting for thermal dissociation effects in terms of chemical by-products and vertical abundances when deriving the atmospheric composition. These challenges are also likely to affect JWST observations, as hinted by \citet{2023arXiv230108192C}. Nonetheless, exoplanet atmospheres are 3D structures, which perhaps makes our 1D forward models insufficient to explain the global chemistry (in particular, for highly-irradiated tidally-locked systems). Therefore, it is necessary to exercise caution when interpreting the retrieved atmospheric properties from 1D retrievals. Furthermore, we fix the model scaling factor, $\alpha$ to $1$ in our model fits and also assume that the atmospheric signal is constant over time. However, the disk-averaged temperature-pressure profile changes as the planet rotates, which might lead the atmospheric models to change with time and/or phase. While the retrieved abundances for WASP-18b were found to be identical (within $1\sigma$) using a model with a fixed scale factor and a phase-dependent scale factor \citep[e.g.][]{brogi2023roasting}, emission spectroscopy analyses of the UHJ, WASP-33b, detected a phase-dependence found via the model scaling parameter and report that a larger scaling is required to best model the observations after the secondary eclipse \citep[e.g.][]{2023MNRAS.522.2145V, 2022AJ....163..248H}. Therefore, temporally parametrizing the scale factor in our forward model and implementing spatial/phase-resolved retrievals can help explore potential variations in composition, temperature, etc., and could also help explain the velocity shifts of the $\rm Fe$ feature. We aim to explore these in future work.

\section{Conclusions}\label{sect6}
In this work, we presented high-resolution emission spectroscopy observations of the ultra-hot Jupiter MASCARA-1b using the upgraded CRIRES+ spectrograph installed at the VLT. We apply the standard cross-correlation methodology  as well as a retrieval analysis and learn the following about the thermal and chemical properties of the planet:
\begin{itemize}
    \item We detected strong emission signatures of $ \mathrm{CO} $ (${\approx}17\sigma $), $ \mathrm{H_2O} $ (${\approx}11\sigma $) and reported the K-band detection of $ \mathrm{Fe} $ (${\approx}8\sigma$) in the day-side atmosphere of MASCARA-1b (Sec.~\ref{sec:4.1}). We also confirm the presence of a temperature inversion layer.\\
    \item The likelihood framework introduced in \citet{gibson2020detection} was applied to obtain quantitative information about the planet's composition. Our retrieval framework also allowed us to constrain the abundances, $T$-$P$ profile, planetary orbital velocity, $\rm C/O$ ratio, as well as metallicity while simultaneously marginalising over the noise properties of the data set.\\
    \item A tentative evidence for shifts in the systemic velocity ($ v_\mathrm{sys} $) and $ K_\mathrm{p} $ is seen for the $ \mathrm{Fe} $ feature, and we advocate for follow-up studies to confirm these shifts (Sec.~\ref{sect5}).\\
    \item In this study, we implemented four different models: two parametric $T$-$P$ profiles (Sec.~\ref{sec:3.2}) and two different chemical regimes (Sec.~\ref{sec:4.2}).\\
    \item We highlight the shortcomings of a free-retrieval model assuming a well-mixed atmosphere. Such a model points to a super-solar $ \mathrm{C/O} $ ratio of $ 0.98^{+0.01}_{-0.02} $. The elevated $\rm CO$ abundance relative to $\rm H_2O$ drives the $\rm C/O$ ratio towards 1 as well as results in an unrealistically small uncertainty, which could lead to biases in UHJs (Sec.~\ref{sec:4.2}).\\
    \item Incorporating a self-consistent chemical model in our retrieval results in a $\rm C/O$ of $0.68^{+0.12}_{-0.22}$ and a metallicity, $\rm [M/H] = 0.62^{+0.28}_{-0.55}$, both consistent with the solar value within $\approx$1.1$\sigma$ (Sec.~\ref{sec:4.2}). Additionally, we tested the effect of adjusting the $\rm C/O$ ratio on our retrievals (i.e. varying C or O or both species) and find that the choice of scaling did not impact the retrieved $\rm C/O$ ratios (Sec.~\ref{sect5}).\\
    \item We tested the effect of \textsc{SysRem} iterations on our retrievals and find that using $N$\,=\,5,\,10 and 15 passes give us consistent results, whereas $N$\,=\,20 filtered out the $\rm Fe$ signal (Sec.~\ref{sec:4.2}).\\
\end{itemize}

While our free-retrieval results strongly favoured a $\rm H_2O$-depleted model over one with a similar abundance of $\rm CO$ and $\rm H_2O$, the non-detection of $\rm OH$ in the day-side atmosphere of MASCARA-1b allows for more precise abundance estimations to be made in the future, which could help identify whether or not the low $\rm H_2O$ abundance is likely the result of thermal dissociation, which has been proposed to be a possibility for ultra-hot Jupiters. Overall, this study is a strong validation of our model filtering and retrieval frameworks, as well as the performance of CRIRES+ for high-resolution emission spectroscopic studies of ultra-hot Jupiters.

\section*{Acknowledgements}
We are extremely grateful to the anonymous referee for careful reading of the manuscript and comments that improved the clarity of the paper. This work relied on the observations collected at the European Organisation for Astronomical Research in the Southern Hemisphere under ESO programme 107.22TQ.001 as part of the CRIRES+ Science Verification run. We are extremely grateful to the CRIRES+ instrument teams and observatory staff who made these observations possible. S.R. gratefully acknowledges support from a Provost's PhD Project Award from Trinity College Dublin. N.P.G and C.M. are supported by Science Foundation Ireland and the Royal Society in the form of a University Research Fellowship and Enhancement Award. S.K.N is supported by JSPS KAKENHI grant No. 22K14092.
We are grateful to the developers of the NumPy, SciPy, Matplotlib, corner, petitRADTRANS, \textsc{FastChem}, and Astropy packages, which were used extensively in this work \citep{harris2020array, virtanen2020scipy, hunter2007matplotlib, perez2007ipython, foreman2016corner, molliere2019petitradtrans, stock2018fastchem}.

\section*{Data Availability}
The observations detailed in this publication are publicly available in the ESO Science Archive Facility (\url{http://archive.eso.org}) under the program name 107.22TQ.001. Data products will be shared on reasonable request to the corresponding author.



\bibliographystyle{mnras}
\bibliography{Bibliography} 




\appendix

\section{Some extra material}\label{appendixA}
We have included additional plots regarding our injection tests detailed in Section~\ref{sec:2.1}, and retrieval frameworks described in Section~\ref{sec:4.2} using a different parametric $T$-$P$ profile below, similar to Figures~\ref{fig:chemmodel_guillot} and~\ref{fig:chem_ret_tpenvelope}. Additional 1D and 2D posterior distributions including potential C- and O-bearing species (i.e. $ \mathrm{CO} $, $\mathrm{H_2O}$, $\mathrm{Fe}$, $\mathrm{CO_2}$, $\mathrm{HCN}$, $\mathrm{OH}$, and $\mathrm{CH_4}$) detailed in Section~\ref{sect5} have also been included. A table of retrieved parameters for the combined fits of MASCARA-1b using the parametric profile from \citet{madhusudhan2009temperature} (similar to Table~\ref{retrievalpars_guillot}) is also given in Table~\ref{retrievalpars_mstp} below.

\begin{figure*}
  \centering
  \includegraphics[width=2.1\columnwidth]{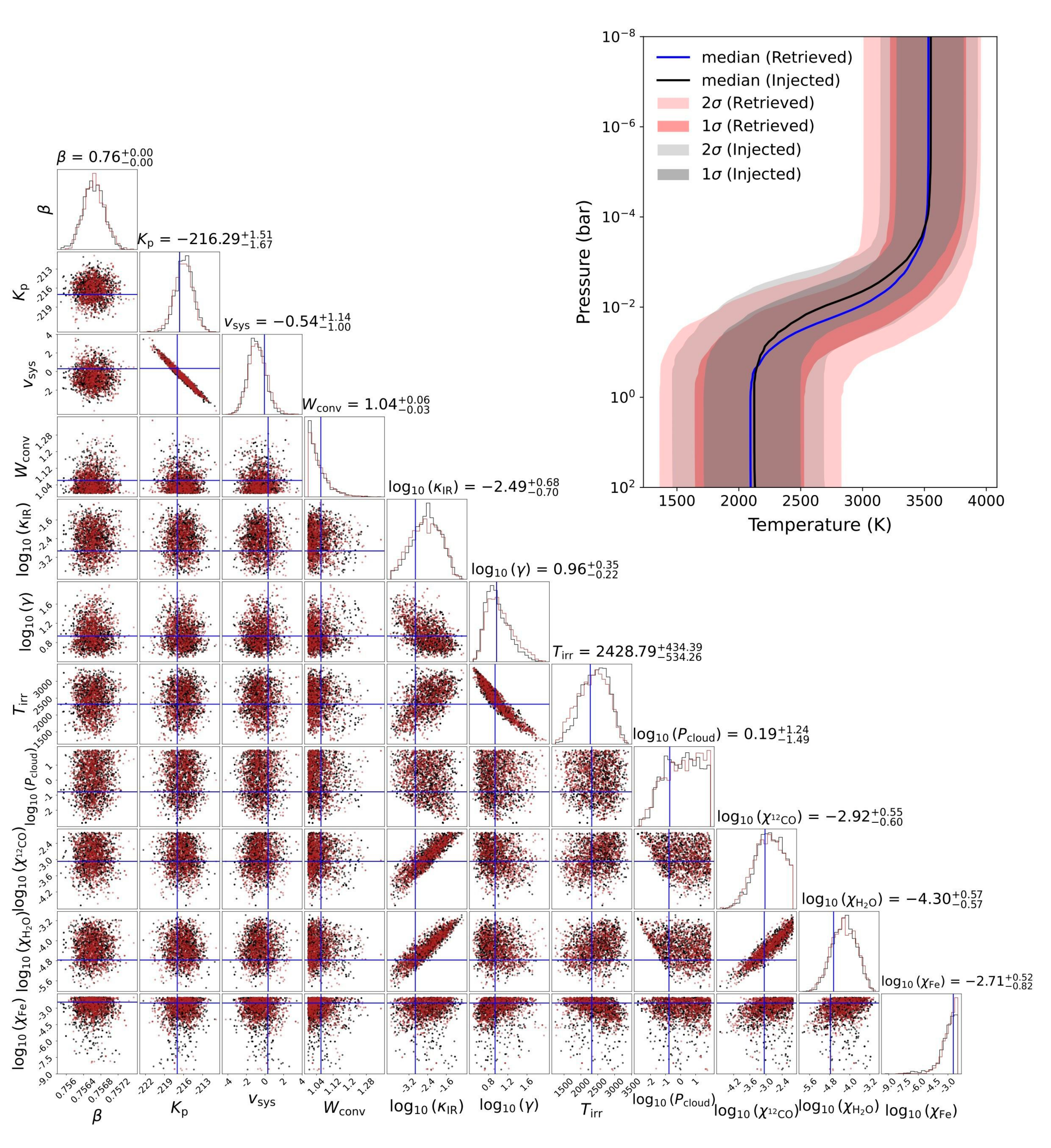}
  \caption{The results of our injection tests described in Section~\ref{sec:2.1} for MASCARA-1b. Left: 1D and 2D marginalised posterior distributions of the parameters from the MCMC fit. The different colours show the samples from two different sets of walkers, with the injected values shown as blue solid lines, respectively. \textit{Upper right:} The retrieved \textrm{$T$-$P$} profile and its injected value are shown as blue and black solid lines. The red and grey shadings mark the 1$\sigma$ and 2$\sigma$ recovered distribution computed from $10{,}000$ samples from the MCMC.}\label{fig:injection_test}
\end{figure*}

\begin{figure*}
  \centering
  \includegraphics[width=2.2\columnwidth]{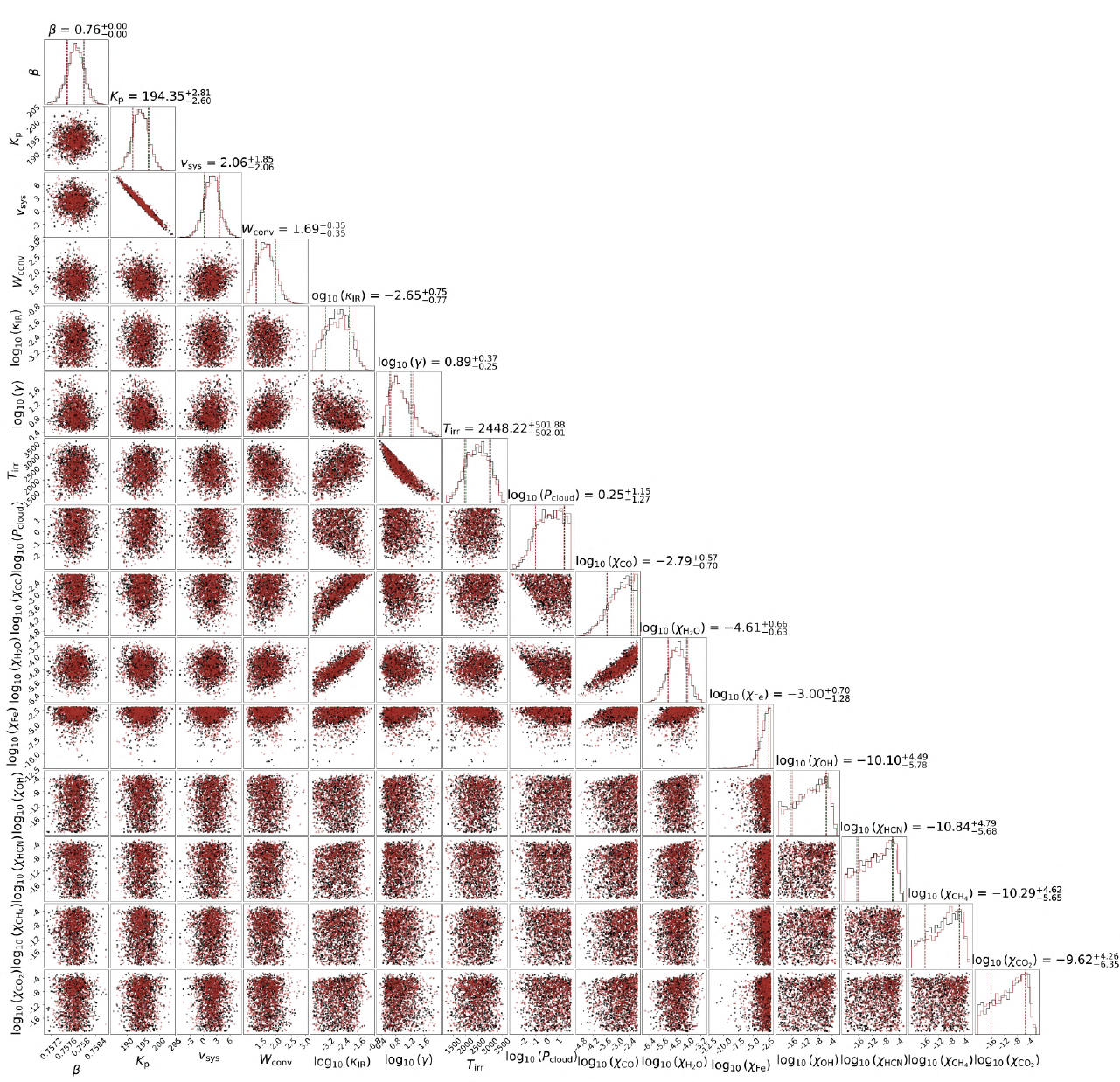}
  \caption{A summary of our free retrieval results with the inclusion of $ \mathrm{OH} $, $\mathrm{CO_2}$, $\mathrm{CH_4}$, and $\mathrm{HCN}$ as described in Section~\ref{sect5} with the 1D and 2D marginalised posterior distributions of each model parameters displayed within a corner plot using the parametric profile from \citet{guillot2010radiative}.}\label{fig:freeret_allspecies}
\end{figure*}

\begin{figure*}
  \centering
  \includegraphics[width=1.97\columnwidth]{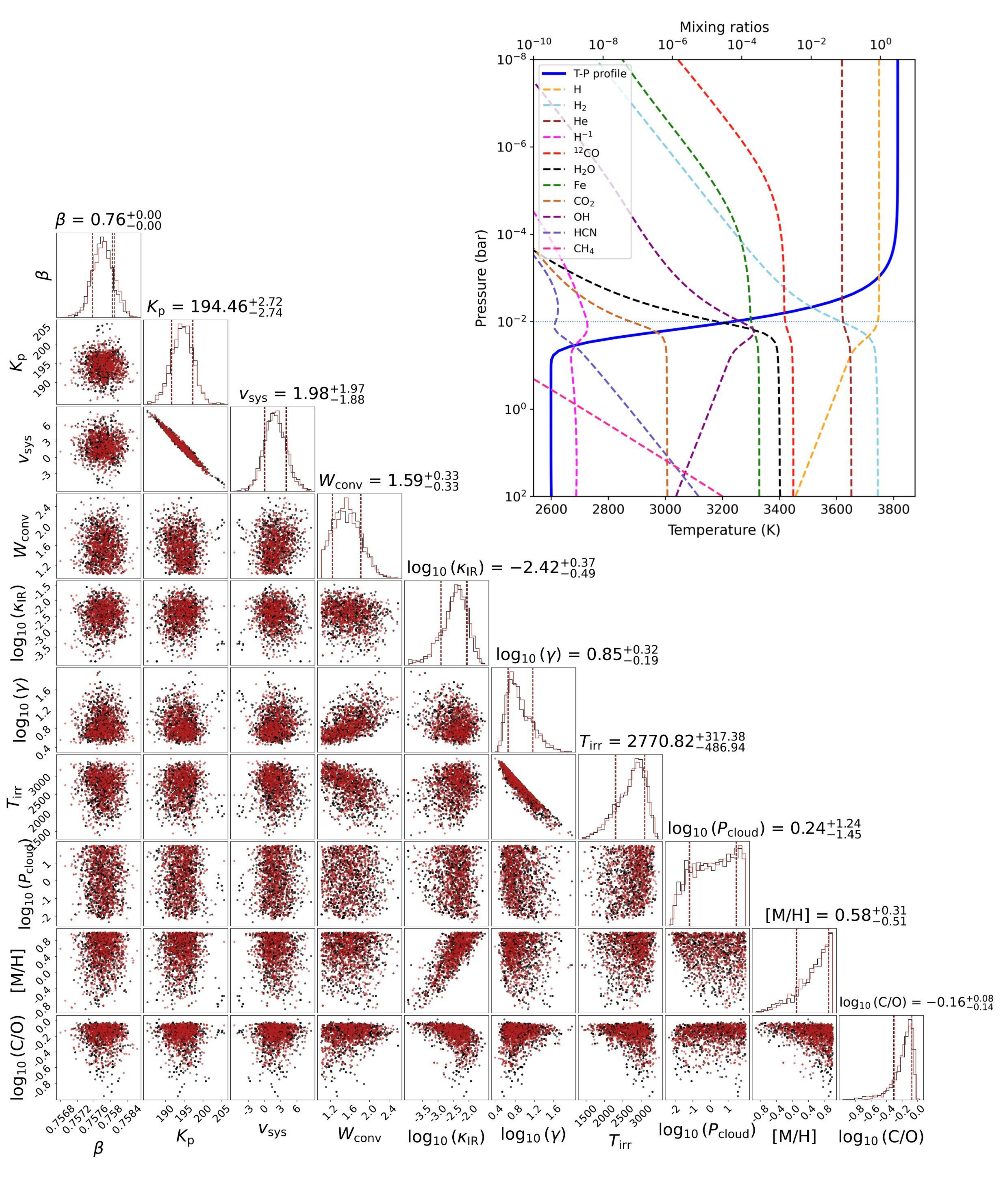}
  \caption{Similar to Fig.~\ref{fig:freeret_allspecies}, but assuming equilibrium chemistry. \textit{Upper right:} the atmospheric structure from the best-fitting chemical model. The volume mixing ratio profiles for the species are shown as dashed lines (calculated using FastChem) and the parametric $T$$-$$P$ profile is shown as a solid blue line.}\label{fig:chemret_allspecies}
\end{figure*}

\begin{figure*}
  \centering
  \includegraphics[width=1.95\columnwidth]{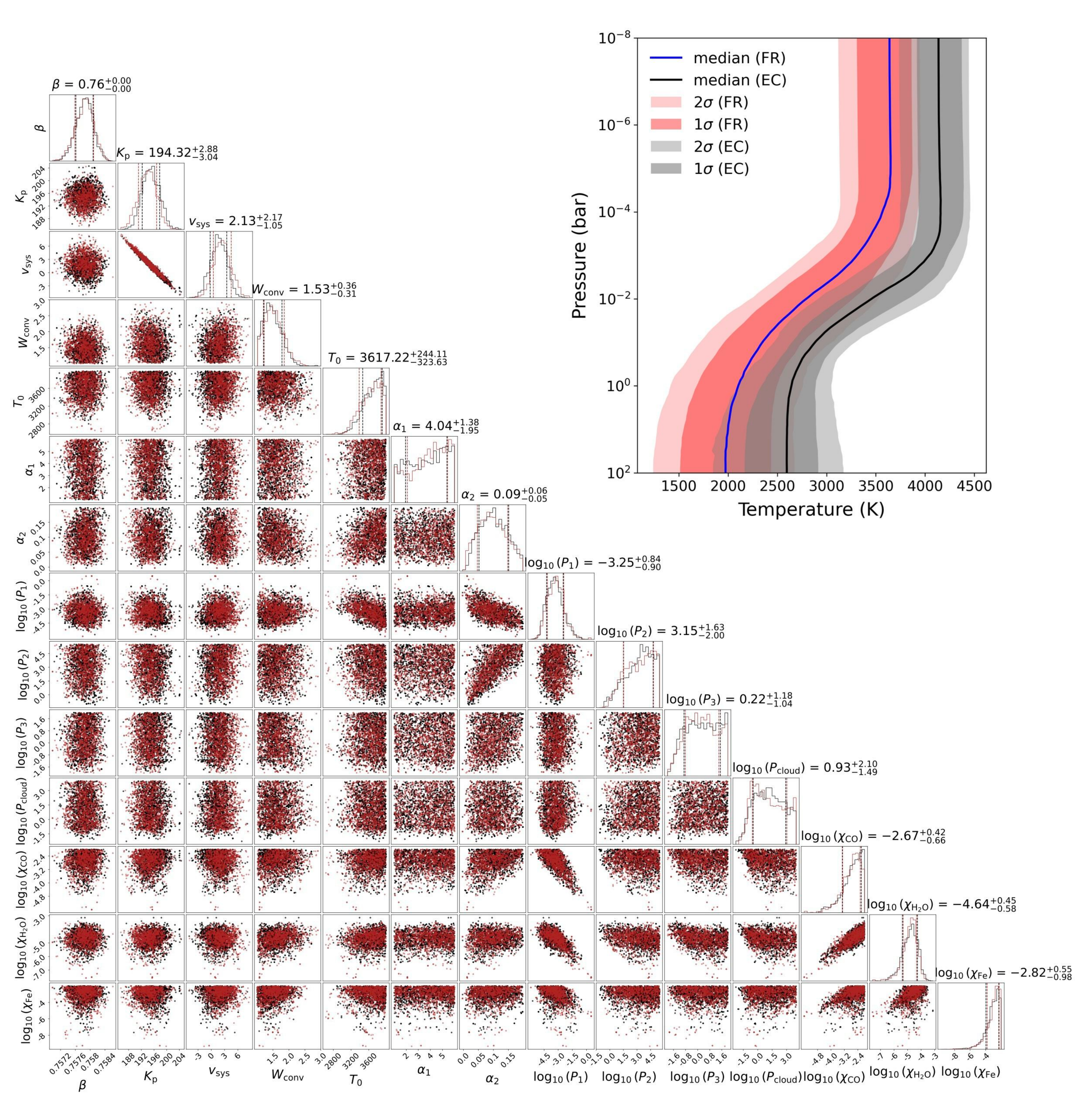}
  \caption{The results for our free-retrieval setup for MASCARA-1b described in Section~\ref{sec:4.2} with the 1D and 2D marginalised posterior distributions of each model parameters using the parametric profile from \citet{madhusudhan2009temperature}. \textit{Upper right:} the $T$$-$$P$ profile computed from the best-fitting model parameters. The black and blue lines are the median profiles for the free-retrieval (labelled FR) and the chemistry (labelled EC) frameworks. The grey and red shading marks the $ 1\sigma $ and $ 2\sigma $ recovered distribution computed from 10,000 samples from the MCMC for both regimes.}\label{fig:chemmodel_mstp}
\end{figure*}

\begin{table*}
\centering
  \caption{Parameters recovered for the combined fits of MASCARA-1b for two different chemical regimes using the parametric $T$-$P$ profile from \citet{madhusudhan2009temperature}.}
  \label{retrievalpars_mstp}
  \renewcommand{\arraystretch}{1.55}
  \begin{tabular}{ccccc}
    \hline
    Parameter [units] & Prior & Free-retrieval & Equilibrium chemistry\\
    \hline
     $ \alpha $ & - & - & - \\
     $ \beta $ & $ \mathcal{U}(0.1, 2) $ & $ 0.76 \pm 0.0003 $ & $ 0.76 \pm 0.0003 $ \\
     $ K_\mathrm{p} $ [$ \rm km $ $\rm s^{-1}$] & $ \mathcal{U}(185, 215) $ & $ 194.8^{+2.9}_{-3.0} $ & $ 194.5^{+2.6}_{-2.5} $ \\
     $ v_\mathrm{sys} $ [$ \rm km $ $\rm s^{-1}$] & $ \mathcal{U}(-15, 15) $ & $ 1.75 \pm 2.00 $ & $ 1.82^{+1.78}_{-1.74} $ \\
     $ W_\mathrm{conv} $ & $ \mathcal{U}(1, 50) $ & $ 1.50^{+0.36}_{-0.29} $ & $ 1.40^{+0.31}_{-0.26} $ \\
     $ T_\mathrm{0} $ [$\mathrm{K} $] & $ \mathcal{U}(1000, 4000) $ & $ 3650^{+230}_{-310} $ & $ 4151^{+227}_{-255} $ \\
     $ \alpha_1 $ & $ \mathcal{U}(1, 6) $ & $ 3.95^{+1.45}_{-1.94} $ & $ 4.41^{+2.37}_{-2.29} $ \\
     $ \alpha_2 $ & $ \mathcal{U}(-1, 0.2) $ & $ 0.10^{+0.05}_{-0.05} $ & $ 0.05^{+0.03}_{-0.03} $ \\
     $ \log_{{10}}(P_1) $ [bar] & $ \mathcal{U}(-5.5, 2.5) $ & $ -3.31^{+0.88}_{-0.88} $ & $ -2.84^{+0.59}_{-0.62} $ \\
     $ \log_{{10}}(P_2) $ [bar] & $ \mathcal{U}(-5.5, 3.5) $ & $ 3.21^{+1.54}_{-2.05} $ & $ 1.70^{+1.24}_{-1.47} $ \\
     $ \log_{{10}}(P_3) $ [bar] & $ \mathcal{U}(-2, 2) $ & $ 0.25^{+1.21}_{-1.12} $ & $ 0.07^{+1.32}_{-1.18} $ \\
     $ \log_{{10}}(P_\mathrm{cl}) $ [bar] & $ \mathcal{U}(-4, 2) $ & $ 1.05^{+1.92}_{-1.57} $ & $ 0.26^{+1.22}_{-1.18} $ \\
     $ \log_{{10}}(\chi_{\rm CO}) $ & $ \mathcal{U}(-20, -2) $ & $ -2.66^{+0.44}_{-0.67} $ & - \\
     $ \log_{{10}}(\chi_{\rm H_2O}) $ & $ \mathcal{U}(-20, -2) $ & $ -4.64^{+0.48}_{-0.61} $ & - \\
     $ \log_{{10}}(\chi_{\rm Fe}) $ & $ \mathcal{U}(-20, -2) $ & $ -2.88^{+0.60}_{-1.00} $ & - \\
     $ [\mathrm{M/H}] $ & $ \mathcal{U}(-1, 1) $ & $ 0.67^{+0.36}_{-0.65} $ & $ 0.67^{+0.24}_{-0.40} $ \\
     $ \log_{{10}}(\mathrm{C/O}) $ & $ \mathcal{U}(-1, 1) $ & $ -0.011^{+0.01}_{-0.00} $ & $ -0.32^{+0.21}_{-0.30} $ \\
    \hline
  \end{tabular}
\end{table*}

\begin{figure*}
  \centering
  \includegraphics[width=0.96\columnwidth]{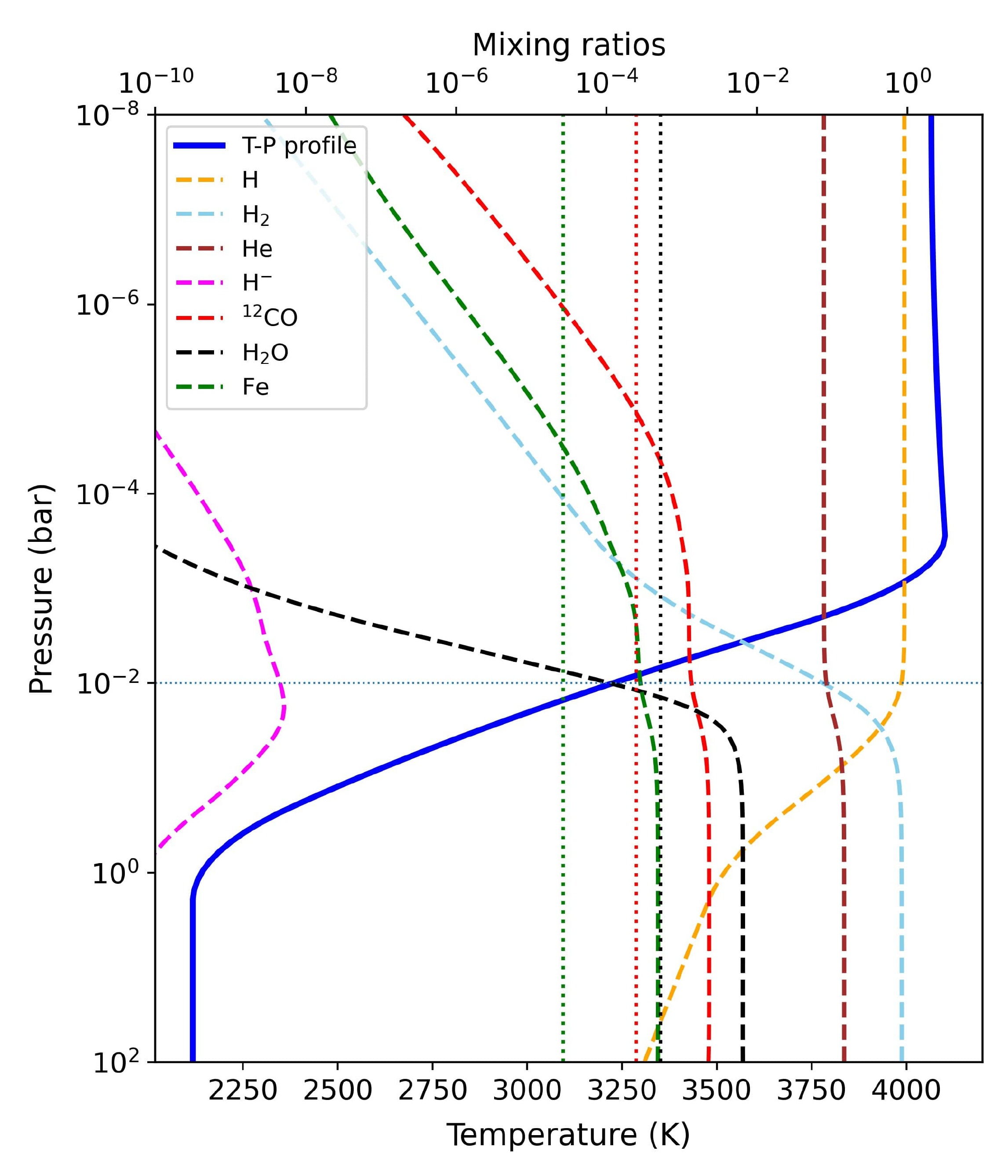}
  \caption{The atmospheric structure from the best-fitting equilibrium chemistry model. The volume mixing ratio profiles for $ \mathrm{CO} $, $ \mathrm{H_2O} $ and $ \mathrm{Fe} $ are shown as as dashed lines (calculated using {\fontfamily{pcr}\selectfont FastChem}). The parametric $T$-$P$ profile \citep{madhusudhan2009temperature} is shown as a solid blue line.}\label{fig:chem_ret_mstp_chemmodel}
\end{figure*}

\begin{figure*}
  \centering
  \includegraphics[width=2.15\columnwidth]{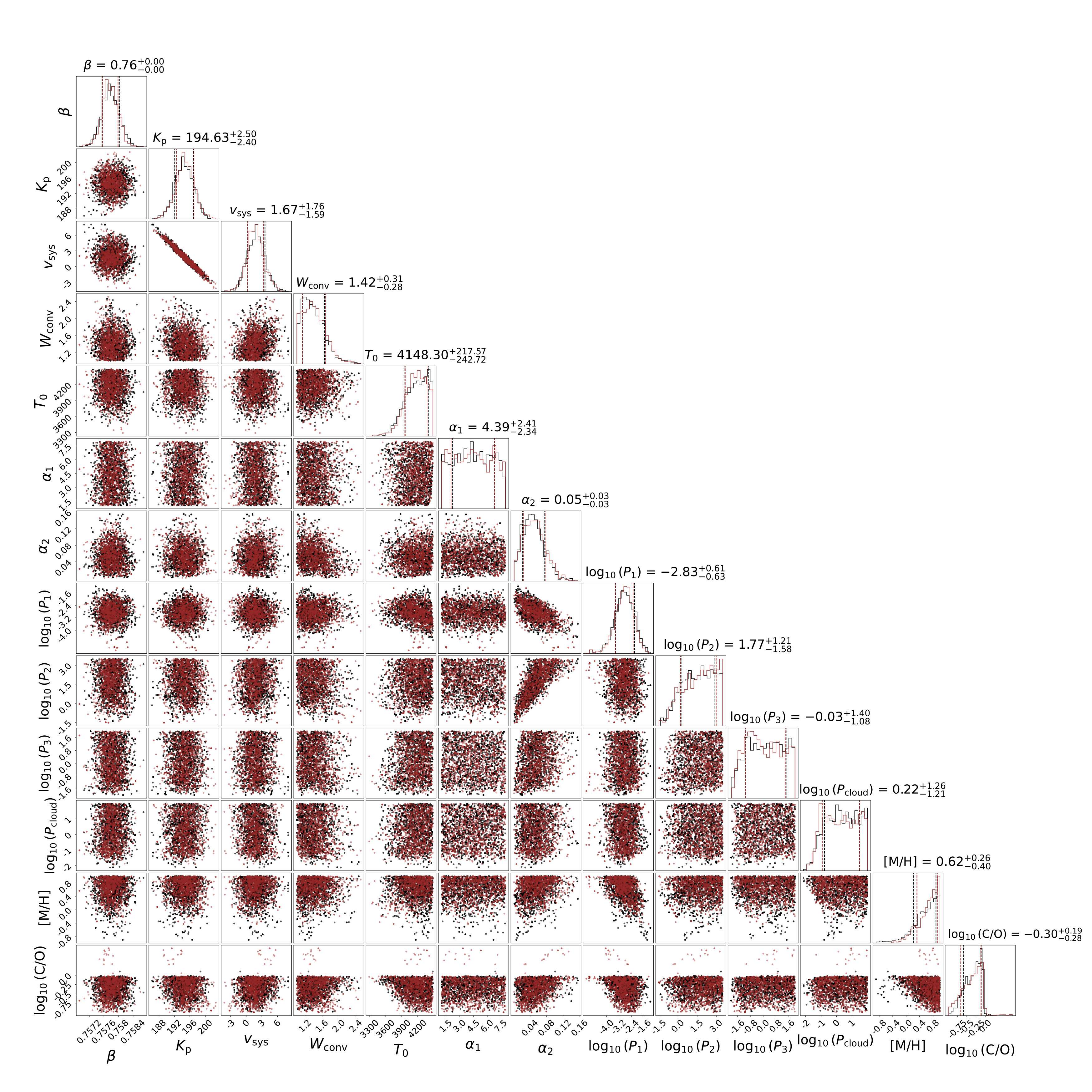}
  \caption{Similar to Fig.~\ref{fig:chemmodel_mstp}, but for an equilibrium chemistry model.}\label{fig:chem_ret_mstp_tpenvelope}
\end{figure*}

\bsp	
\label{lastpage}
\end{document}